\documentclass[sigplan,screen,10pt]{acmart}

\AtBeginDocument{%
  }

\usepackage{multirow}
\usepackage{booktabs}
\usepackage{graphicx}
\usepackage{braket}
\usepackage{subcaption}
\usepackage{amsmath}

\usepackage{amssymb}
\usepackage{xspace}
\usepackage{balance}
\usepackage[inline]{enumitem}
\usepackage[table]{xcolor} 
\setlist{nosep}
\usepackage{pifont}
\usepackage[most]{tcolorbox}

\definecolor{headergray}{RGB}{210,210,210}
\definecolor{bluegray}{rgb}{0.4, 0.6, 0.8}
\definecolor{pastelblue}{rgb}{0.68, 0.78, 0.81}
\definecolor{palecornflowerblue}{rgb}{0.67, 0.8, 0.94}
\definecolor{paleaqua}{rgb}{0.74, 0.83, 0.9}
\definecolor{lightcornflowerblue}{rgb}{0.6, 0.81, 0.93}
\definecolor{lightblue}{rgb}{0.68, 0.85, 0.9}
\definecolor{cambridgeblue}{rgb}{0.64, 0.76, 0.68}
\definecolor{ashgrey}{rgb}{0.7, 0.75, 0.71}
\definecolor{beaublue}{rgb}{0.74, 0.83, 0.9}
\definecolor{lightcornflowerblue}{rgb}{0.6, 0.81, 0.93}
\definecolor{moonstoneblue}{rgb}{0.45, 0.66, 0.76}
\definecolor{palecornflowerblue}{rgb}{0.67, 0.8, 0.94}
\definecolor{powderblue}{rgb}{0.69, 0.88, 0.9}
\definecolor{blockblue}{RGB}{220,230,242}
\definecolor{blockgreen}{RGB}{221,235,229}
\definecolor{blockyellow}{RGB}{240,232,210}

\definecolor{lightblue}{HTML}{E6F0FF}
\definecolor{lightgreen}{HTML}{E9F7EF}
\definecolor{lightyellow}{HTML}{FFF9E6}
\definecolor{lightpink}{HTML}{FDEDEC}
\definecolor{headergray}{HTML}{F2F2F2} 

\definecolor{headerteal}{RGB}{100,180,180}
\definecolor{lightyellow}{RGB}{255,255,180}
\definecolor{lightlavender}{RGB}{210,190,240}

\definecolor{headerlight}{RGB}{180,220,220}

\definecolor{lightyellow}{HTML}{E6F1FB}
\definecolor{lightlavender}{HTML}{FAEEDA}
\definecolor{lightlavendernew}{HTML}{EDE9FE}

\definecolor{itergray}{RGB}{220, 220, 220}
\definecolor{mapone}{RGB}{198, 235, 210}   
\definecolor{maptwo}{RGB}{198, 218, 240}   
\definecolor{mapthree}{RGB}{255, 223, 186} 

\newtcolorbox{observationbox}[1][]{
  enhanced,
  colback=white,
  colframe=gray!50,
  colbacktitle=gray!30,
  coltitle=black,
  fonttitle=\itshape\bfseries,
  title=#1,
  boxrule=0.5pt,
  arc=2mm,
  left=3mm, right=3mm, top=2mm, bottom=2mm,
  fontupper=\itshape,
}

\usepackage{xcolor}           
\usepackage[normalem]{ulem}   
\usepackage{array}

\newif\ifshowmarkup
\showmarkuptrue

\newcommand{\add}[1]{\ifshowmarkup\textcolor{blue}{#1}\else#1\fi}

\makeatletter
\newcommand{\del@cite}{\@ifnextchar[{\del@citeopt}{\del@citenoopt}}
\def\del@citeopt[#1]#2{\mbox{\del@origcite[#1]{#2}}}
\def\del@citenoopt#1{\mbox{\del@origcite{#1}}}
\def\del@ref#1{\mbox{\del@origref{#1}}}
\newcommand{\del}[1]{%
  \ifshowmarkup
    {\color{red}\bfseries        
     \let\del@origcite\cite \let\cite\del@cite
     \let\del@origref\ref   \let\ref\del@ref
     \sout{#1}}%
  \fi}
\makeatother
\renewcommand{\add}[1]{#1}

\author{Mohammad Abrarul Hasanat}
\affiliation{%
  \institution{University of Utah}
  \city{Salt Lake City}
  \country{USA}
}

\author{Amit Samanta}
\affiliation{%
  \institution{University of Utah}
  \city{Salt Lake City}
  \country{USA}
}

\author{Jason Ludmir}
\affiliation{%
  \institution{Rice University}
  \city{Houston}
  \country{USA}
}

\author{Tirthak Patel}
\affiliation{%
  \institution{Rice University}
  \city{Houston}
  \country{USA}
}

\author{Rohan Basu Roy}
\affiliation{%
  \institution{University of Utah}
  \city{Salt Lake City}
  \country{USA}
}

\renewcommand{\shortauthors}{Hasanat et al.}
\newcommand{\sol}{EntangledEdge\xspace}

\copyrightyear{2026}
\acmYear{2026}
\setcopyright{cc}
\setcctype{by}
\acmConference[SIGOPS ATC 2026]{The 2026 ACM SIGOPS Annual Technical Conference}{November 15-18, 2026}{Hong Kong, China}
\acmBooktitle{The 2026 ACM SIGOPS Annual Technical Conference (SIGOPS ATC 2026), November 15-18, 2026, Hong Kong, China}
\acmDOI{10.1145/3828523.3857119}
\acmISBN{979-8-4007-2818-1/2026/11}

\let\origcopyrightbox\footnotetextcopyrightpermission
\renewcommand\footnotetextcopyrightpermission[1]{%
  \origcopyrightbox{This paper is accepted at the 2026 ACM SIGOPS Annual Technical Conference (SIGOPS ATC 2026), November 15--18, 2026, Hong Kong, China.}}
  
\begin{document}
\title{EntangledEdge: A Distributed System for Orbital
Quantum Sensing and Learning}

\begin{abstract}
\textit{An orbital quantum sensor is more precise than classical and holds its information in a quantum state's superposition and entanglement, which classical measurement collapses. Learning on quantum-sensed data \add{therefore requires a distributed system} in which the data stay quantum from sensor to inference. We present \sol{}, the first \add{system architecture} to unify orbital quantum sensing and learning end-to-end, with a quantum data plane and a classical control plane. It routes each event to a channel-matched model in a capacity-diverse bank, adapts its detection threshold without labels, and aggregates across stations by hardware fidelity.
\add{Evaluated with a hybrid prototype (real quantum processors, simulated orbit-to-ground link) over satellite telemetry, \sol{} achieves high-quality inference at bounded per-event cost, improves quantum processor utilization, and scales with fleet size.} }
\end{abstract}



\setlength{\abovedisplayskip}{4pt}
\setlength{\belowdisplayskip}{4pt}
\setlength{\abovedisplayshortskip}{2pt}
\setlength{\belowdisplayshortskip}{2pt}

\hypersetup{
  colorlinks=true,
  linkcolor=black,
  citecolor=black,
  urlcolor=black,
  filecolor=black,
}

\maketitle

\vspace{-2mm}
\section{Introduction}
\label{sec:intro}

\noindent\textbf{Quantum sensing in orbit.}
A quantum sensor uses a qubit probe to interact with a physical field and records the result in its quantum state. For a growing set of space-sensing tasks, the information of interest is held in the quantum structure of that state, in its superposition and entanglement, which a classical measurement collapses before it can be used. For example, entanglement-enhanced gradiometry and networked clock comparison are two such tasks~\cite{wu2023quantum,
yang2025clock}. A learner with coherent access to the states needs exponentially fewer samples than any classical learner working from measured values~\cite{huang2022quantum,
huang2021power}, so these types of tasks can benefit from an end-to-end quantum pipeline, from the orbital probe through to ground processing.

Quantum sensors also exceed classical sensitivity. A qubit probe reaches sensitivities orders of magnitude beyond fluxgate, mechanical, and oscillator-based instruments. Atomic magnetometers, for instance, measure at the femtotesla scale, while fluxgate instruments operate near the picotesla scale. An atomic-transition reference also removes the periodic recalibration classical space instruments require~\cite{RevModPhys.89.035002, jebraeilli2026stqs}. This is why quantum sensors \add{can be used} for space tasks like geomagnetic field mapping, space-weather characterization, and gravity-anomaly tracking for ice-mass and hydrology~\cite{RevModPhys.89.035002}.

\vspace{1.5mm}

\noindent\textbf{\add{Recent hardware makes an orbital quantum system feasible. }} 
\add{Orbital quantum sensors have flown as first demonstrations, with classical readout~\cite{williams2024pathfinder, li2025realization}. Neutral atom platforms are commercially available and are good candidates for deployed sensors. }
Because the sensed quantum state has to reach the ground intact, the orbit-to-ground link has to be quantum, which teleportation over a pre-shared entangled pair provides. Quantum networking has reached multi-node operation over fiber, and this transfer works between non-neighboring nodes~\cite{pompili2021multi, hermans2022qubit} \add{and at Hertz rates across a city~\cite{shen2023hertz}}. In orbit, \add{the link has been demonstrated as a proof of concept.} The Micius mission distributed entanglement to two ground sites 1200~km apart and demonstrated ground-to-satellite teleportation of a single-photon qubit~\cite{lu2022micius, yin2017satellite, ren2017ground}, and a 23-kg microsatellite payload reproduced satellite-to-ground transmission with portable ground stations~\cite{li2025microsatellite}. On the ground, deployed quantum processors with tens to hundreds of qubits can run the learning and inference~\cite{wurtz2023aquila}. \add{\sol{} is a system architecture for this chain, whose stages have been demonstrated separately.}

\vspace{1.5mm}

\begin{figure}[t]
  \centering
  \includegraphics[width=0.99\columnwidth]{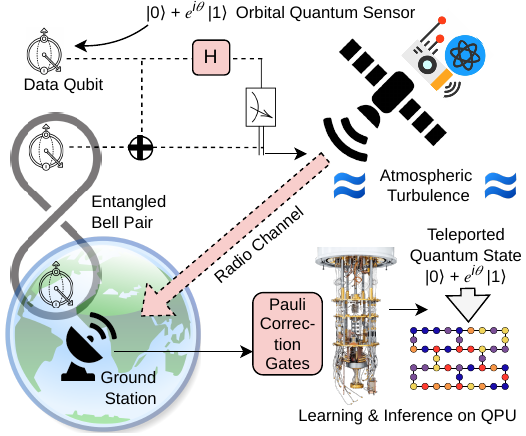}
  \vspace{-4mm}
  \caption{\textit{An orbital quantum sensor teleports its data qubit to a ground station's QPUs through a pre-shared Bell pair and a classical radio channel, where an ansatz performs learning and inference on the recovered state.}}
  \vspace{-5mm}
  \label{fig:overview}
\end{figure}

\noindent\textbf{Learning on orbital quantum-sensed data as a distributed system.}
These parts form an end-to-end pipeline that learns and infers on the quantum-sensed states. Orbital edge computing established satellite constellations as distributed computers with a classical data plane~\cite{denby2020orbital,denby2023kodan,liu2025spaceexit, cheng2024eagleeye}, and a quantum data plane raises challenges \add{that classical remedies do not solve} \add{(Table~\ref{tab:positioning} in Sec.~\ref{sec:related} positions \sol{} against existing classes)}. Sensed states cannot be buffered or copied, since no-cloning forbids it and coherence times are far shorter than the gaps between ground-station passes~\cite{buvzek1996quantum}, so a state that reaches a busy receiver is lost. The atmospheric channel distorts each downlinked state by a different amount per event, and the ground processors are physically distinct with noise that drifts between recalibrations. This variation shifts the score distribution over time, and labels for learning cannot be obtained without destroying the state, so the system has to track it without them.

\vspace{1.5mm}

\noindent\textbf{The \sol{} system design.}
\sol{} addresses these challenges in one end-to-end design with a quantum data plane and a classical control plane. An orbital sensor teleports each state to a ground station, which recovers and processes it, and the stations share only classical model updates (Fig.~\ref{fig:overview}). Because the channel changes per event, the system keeps several models of different capacities and routes each state to the one whose channel suits. It sets its own detection cutoff and updates it as the scores shift, so it needs no labels. Across ground stations, each weights the others by how reliably they recover, so drift and mismatch between processors do not skew the combined system. 

\vspace{1mm}

\textbf{Overall, we make the following contributions:}

\vspace{1mm}

\noindent\textbf{1. \add{Problem and system architecture.}} \add{We formulate learning on orbital quantum-sensed data as a distributed-systems problem in which the data cannot be stored, copied, or re-sent, and present \sol{}, an end-to-end system architecture for orbital quantum sensing, learning, and inference.}

\vspace{1mm}

\noindent\textbf{2. Design to address quantum-specific challenges.} To address the per-event channel and the unlabeled, drifting stream, \sol{} introduces a channel-conditioned multi-model bank, a label-free dynamic threshold, and a fidelity-weighted aggregation, which together add robustness to drifting processor/channel noise and station outages.

\vspace{1mm}

\noindent\textbf{3. \add{Hybrid-prototype evaluation.}} We build and evaluate \sol{} on superconducting processors (IBM Heron and Nighthawk), neutral atom sensors (QuEra Aquila), and \add{a simulated orbit-to-ground link,} over satellite-telemetry.

\vspace{1mm}

\noindent\textbf{4. Quantum model for detection.} We develop autoencoder methods to operate on sensed data and show they improve inference quality over a state-of-the-art quantum autoencoder (Quorum~\cite{ludmir2025quorum}) and federated alternatives in our setting.

\vspace{1mm}

\noindent\textbf{5. Systems performance and scaling.} \sol{} delivers its detection gains without hurting latency or throughput, improves processor utilization compared to alternatives, is robust across different operating conditions, and scales across sensors, ground stations, and model sizes.

\vspace{1mm}
\sol{}'s implementation is open-sourced at \newline \href{https://doi.org/10.5281/zenodo.22909579}{https://doi.org/10.5281/zenodo.22909579}. We hope it helps others to reproduce our results and build on this work.
\vspace{-2mm}
\section{Background and Motivation} 
\label{sec:bg}


\noindent A satellite-based quantum sensing pipeline has two components.
An \emph{orbital sensor node} is a satellite payload carrying qubits that interact with a physical field. A \emph{ground QPU} is an optical ground station that decodes the sensed state and processes it on a quantum processing unit (QPU).
The two are connected by a quantum link and a classical radio link.
Similar partitioning of work between satellite and ground stations appears in orbital edge computing for classical compute-based satellite payloads~\cite{denby2020orbital,
denby2023kodan,
10121575}.

\vspace{-2mm}
\subsection{Qubits, gates, and quantum machine learning}
\label{sec:bg:qc}
A qubit is a two-level quantum system that can be in a superposition of its basis states $\ket{0}$ and $\ket{1}$.
Two or more qubits can occupy \emph{entangled} states. The simplest example is
$\tfrac{1}{\sqrt{2}}(\ket{00}+\ket{11})$, in which measuring either
qubit fixes the outcome of the other.
Operations on qubits are unitary gates, single-qubit
rotations $R_x(\theta), R_y(\theta), R_z(\theta)$,
and the two-qubit
entangling controlled-NOT (CNOT), applied as a circuit that ends in measurement.
A Quantum Machine Learning (QML) model is a parameterized circuit $U(\boldsymbol{\theta})$, an \emph{ansatz}, whose gate angles form a
trainable vector $\boldsymbol{\theta}$. Training proceeds in a hybrid
quantum-classical loop. Input data, delivered directly to the
processor, is loaded into a register. The ansatz
$U(\boldsymbol{\theta})$ acts on the register, and
observables are measured to produce a scalar cost
$C(\boldsymbol{\theta})$. A classical
optimizer updates the parameters along the negative gradient
$-\nabla_{\boldsymbol{\theta}} C$, with each partial derivative
estimated from two runs of the same circuit at shifted angles, the parameter-shift rule~\cite{schuld2019evaluating}.

\vspace{-2mm}
\subsection{Quantum sensing}
\label{sec:bg:sensing}
A quantum sensor uses a qubit as a probe, prepared in the superposition
$\tfrac{1}{\sqrt{2}}(\ket{0}+\ket{1})$ and exposed to the field for
a free-evolution time $t$. The field shifts the relative phase
between the $\ket{0}$ and $\ket{1}$ components at a rate
proportional to its strength, and the qubit ends in
$\tfrac{1}{\sqrt{2}}(\ket{0}+e^{i\theta}\ket{1})$, where the
accumulated phase $\theta$ records the integrated field~\cite{RevModPhys.89.035002}. Entangling $N$ probes before exposure improves the error scaling
from the classical $1/\sqrt{N}$ to the Heisenberg limit $1/N$~\cite{giovannetti2006quantum,jebraeilli2026stqs}.
A long coherence time $T_2$ lets the probe
integrate signal, so
sensitivity scales roughly as $1/\sqrt{T_2}$.
Recent work has improved both for sensors of different modalities~\cite{RevModPhys.89.035002,jebraeilli2026stqs, ludmir2026spintune}. 

\vspace{1.5mm}

\noindent\textbf{Orbital quantum sensors.} Orbital quantum sensors open a class of
Earth-observation and space sensing tasks that classical sensors cannot reach.
Quantum probes operate near the shot-noise floor and reach magnetic, gravitational, and frequency sensitivities orders of magnitude beyond fluxgate, mechanical, or oscillator-based instruments. In the setup of Fig.~\ref{fig:overview}, each sensed qubit is
teleported from orbit to a ground QPU. When sensor and ground QPU modalities differ,
a transducer maps the sensed state onto a photonic flying
qubit. Entanglement between optical and microwave
photonic modes has been demonstrated on integrated
hardware~\cite{meesala2024quantum,meesala2024non}.



\vspace{-2mm}
\subsection{Networking via quantum teleportation}
\label{sec:bg:downlink}

Teleportation moves a quantum state across the link without sending
the qubit itself. The two parties pre-share an entangled qubit pair
(a \emph{Bell pair}).
To transfer a data qubit, the sender performs a
Bell-basis measurement on it together with its half of the
Bell pair; the entanglement projects the receiver's qubit
into a Pauli-rotated copy of the original. The sender
transmits the two-bit outcome over an ordinary radio link,
and the receiver applies the matching Pauli correction to recover
the data qubit.
The data qubit itself never crosses the optical channel,
making teleportation the elementary primitive for
quantum networking~\cite{wehner2018quantum}.

Teleportation \add{has been demonstrated} on quantum hardware over fiber and between ground and orbit, so the satellite-ground topology this paper relies on \add{already exists as a proof of concept}~\cite{hermans2022qubit, lu2022micius, yin2017satellite, ren2017ground, li2025microsatellite}. Atmospheric turbulence distorts the optical phase and
spatial mode of each downlinked photon, and the receiving telescope
sees a polarization frame~\cite{han2020polarization}.
At the qubit level, these effects
show up as random single-qubit errors on each teleported qubit, and
the errors grow with the zenith angle and the slant-path
length. 

\vspace{1.5mm}

\noindent\textbf{Design choices.} A design choice is where to
place the QPU. Hosting the QPU on
the satellite would be ideal but is not yet
practical at small-satellite scale. A quantum processor with the
gate depths and fidelities a learning workload needs depends on
infrastructure that does not fit on board today.
Sensor payloads sit at the more permissive end of this
spectrum. We therefore place the QPU at the
ground station for the rest of this paper.
\add{Placing the QPU in orbit instead would remove the satellite-to-ground link and its atmospheric noise, although a quantum link would still remain wherever a sensor and its processor are apart.}

Given that the link is needed, the next question is whether it
needs to carry quantum information at all.
The alternative is to measure the sensed state $\rho$ at the satellite and ship classical outcomes down, which suffices for point estimation but collapses the superposition and entanglement a learner uses to tell states apart. Quantum learners with coherent access reach exponentially lower sample complexity than any classical learner~\cite{huang2022quantum,huang2021power, ludmir2026qufoundrygeneratingdataquantum}, so the quantum link carries strictly more information.
The workload for the rest of this section is anomaly detection on satellite
telemetry, where streams are large, and labeled
events are rare, which is exactly the unsupervised regime in which
this advantage shows up.

\vspace{1.5mm}

\begin{figure}[t]
  \centering
  \includegraphics[width=0.96\columnwidth]{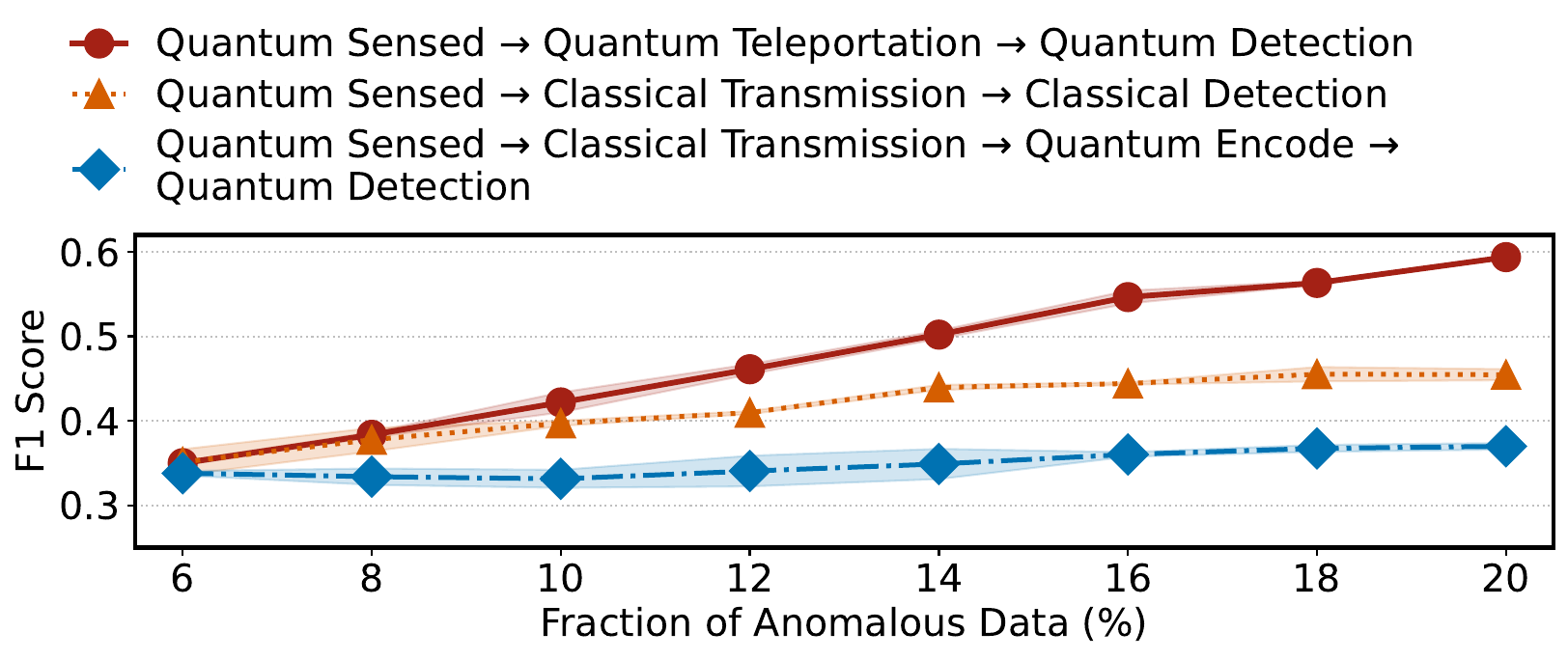}
  \vspace{-4mm}
  \caption{\textit{The fully quantum pipeline leads. The others measure the state at the satellite and lose the quantum structure. Shaded bands are across-seed standard deviations.}}
  \vspace{-4mm}
  \label{fig:motivation_f1}
\end{figure}

\noindent\textbf{Empirical evidence.} Fig.~\ref{fig:motivation_f1} confirms this on OPS-SAT \cite{ruszczak2025ops} telemetry data as
the anomaly fraction sweeps from 6\% to 20\%.
All three pipelines share the same
bottleneck and training budget.
The fully quantum pipeline (sense $\to$ teleport $\to$ quantum autoencoder detection)
holds the highest F1 at every operating point. The
classical-transmission pipeline measures quantum state $\rho$ at the satellite
into a 25-dimensional vector of single-qubit X/Y/Z and two-qubit ZZ
observables, transmits it, and runs a classical autoencoder
on the ground. The hybrid pipeline takes the same satellite-side
measurement and transmission, then re-encodes the feature vector
into a quantum state on the ground,
and runs the same quantum autoencoder. Both
alternatives lose because the satellite-side measurement collapses
$\rho$, destroying the superposition and entanglement structure that
separates anomalous from normal states.
An all-classical baseline
(classical sensor on the satellite, classical processing on the
ground) trails all three pipelines by more than 15\% F1 at every
operating point. That gap reflects the sensitivity headroom quantum
sensors carry at these low signal levels (Sec.~\ref{sec:bg:sensing}).

\vspace{-2mm}
\subsection{Coordination across ground stations}
\label{sec:bg:coordination}

\begin{figure}[t]
  \centering
  \includegraphics[width=0.96\columnwidth]{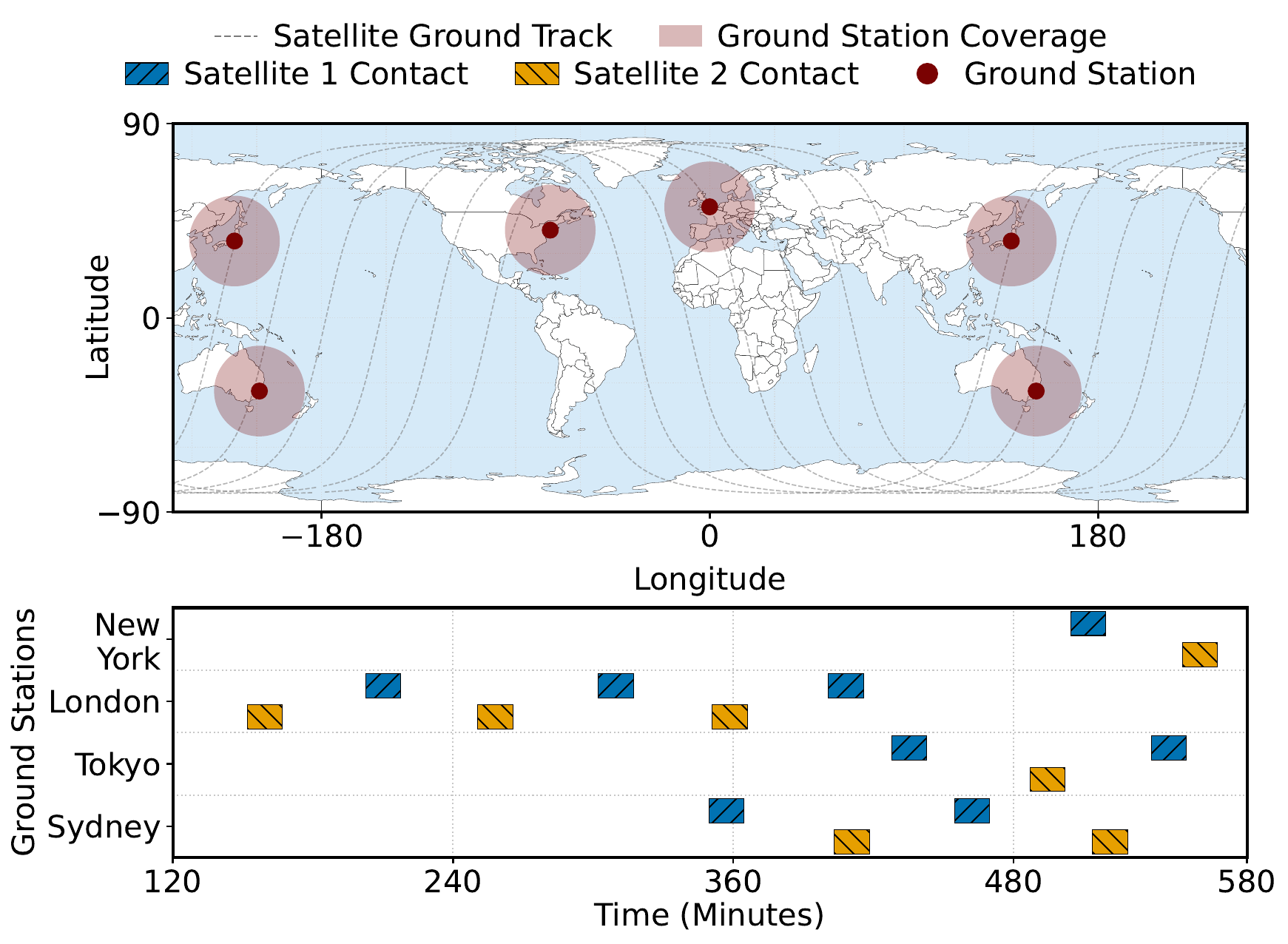}
  \vspace{-4mm}
\caption{\textit{Ground tracks of two satellites (NOAA-20,
NOAA-21), ground-station coverage disks, and 
contact windows. Passes are short, sparse, and not simultaneous
across stations.}}
  \vspace{-5mm}
  \label{fig:coverage}
\end{figure}

\noindent\textbf{Coverage and ground tracks.}
A real deployment runs over many satellites and ground stations,
and the orbital geometry constrains when each satellite-ground link
is available. A station communicates with a satellite only while the
satellite is above its local horizon. As the satellite orbits, the
curve of points directly beneath it (its \emph{ground
track}) sweeps the surface, and a station sees the satellite only
when the ground track passes through its coverage disk.
Fig.~\ref{fig:coverage} (top) shows the shared ground track of
two satellites in the same near-polar orbital plane, with four
representative ground stations
drawn at their 21-degree-elevation coverage disks. The track is
global; each station's coverage disk is a small patch.

\vspace{1mm}

\noindent\textbf{Temporal fragmentation across stations.}
The bottom row of Fig.~\ref{fig:coverage} unrolls these contacts
in time. Each row corresponds to one ground station over a 580-minute
observation window, and the bars within a row mark the intervals
during which each of the two satellites is above that station's
horizon. The pattern drives the design: contact windows at any station are short and sparsely
scattered.
A pass over New York rarely
coincides with a pass over Tokyo, and at any one instant the
satellite delivers data to at most one station.

A natural workaround would be to buffer the sensed states on the
satellite and downlink them later, but this is not an option for
quantum data. Qubit coherence times $T_2$ are on the order of
milliseconds to seconds in the best laboratory quantum memories,
orders of magnitude shorter than the gaps between contact windows. Sensing
and transmission therefore happen in the same brief window, and a
learning workload that depends on the full distribution of sensed
states has to combine what multiple stations see.

\vspace{1.5mm}


\noindent\textbf{Comparison of coordination architectures.}
Three natural coordination architectures fit this setting.
\emph{Isolated} trains a separate model at each ground station on
that station's own data, with no exchange between stations.
\emph{Centralized} routes the sensed data from every station to a
single server and trains one model on the pooled union.
\emph{Federated} keeps each station's data local, trains a model at
each station on its own shard, and exchanges only classical model
parameters between stations for aggregation. We use \emph{federated} here as a label for the distributed
architecture. 
The design choice is driven by communication cost and the noise budget of each architecture's data path, not by data
privacy or security.

Table~\ref{tab:fed_compare} compares the three architectures on a smaller OPS-SAT subset anomaly-detection~\cite{ruszczak2025ops} with a single quantum autoencoder; Sec.~\ref{sec:eval} evaluates the full dataset, averaged over 100 calibration days.
Federated reaches the highest F1. Isolated trails because
each station's local shard is small and biased toward the orbital
patches it sees, so the model overfits to a slice of the
trajectory. Centralized trails for the opposite reason. Pooling routes each station's quantum
state onward over a second ground-to-central link, and the extra
hop adds atmospheric noise to the training data on top of the
satellite-to-ground noise the others pay.
Federated avoids both failure modes: like isolated, it pays only
the satellite-to-ground link, and like centralized each station's model
sees data from across all stations.
The cross-station sharing happens by
classical parameter updates.
Isolated's recall lands a few points above federated's because its
overfitted shard flags more candidates as anomalies, but the same
overfitting collapses its precision and leaves F1 (the right metric for sparse 
anomalies) and accuracy well below federated.
\vspace{1mm}


\begin{observationbox}[Summary of observations]
Orbital quantum sensors reach precision and calibration stability
classical instruments cannot. The sensed state has to remain quantum
end to end, since measuring at the satellite collapses the
state. Ground-station
contacts are short, fragmented across stations, and unbuffered
(decoherence time is orders of magnitude shorter than contact-window), so
a federated architecture with a single quantum link per station and
cross-station sharing of classical model parameters offers the most
benefit.
\end{observationbox}

\vspace{-2mm}
\section{Design of \sol{}} 
\label{sec:des}

\begin{table}[t]
  \centering
  \caption{\textit{Anomaly-detection performance of three coordination
architectures on a smaller OPS-SAT subset; the complete dataset under full per-event noise is in Sec.~\ref{sec:eval}.}}
  \vspace{-4mm}
  \begin{tabular}{lcccc}
    \hline
    \textbf{Architecture} & \textbf{F1} & \textbf{Precision} & \textbf{Recall} & \textbf{Accuracy} \\
    \hline
    Isolated     & 0.51 & 0.39 & 0.76 & 0.70 \\
    Centralized  & 0.44 & 0.30 & 0.82 & 0.58 \\
    Federated    & 0.54 & 0.45 & 0.70 & 0.77 \\
    \hline
  \end{tabular}
  \vspace{-4mm}
  \label{tab:fed_compare}
\end{table}

\noindent \sol{} operationalizes the observations of Sec.~\ref{sec:bg} into
a distributed system. We define an event as one teleported
quantum state received at a ground station, together with the
station's telemetry at the time of reception. In our design, the data plane is
quantum end-to-end, the control plane is classical, and channel
variability, hardware drift, and event loss are first-order
design constraints.

At one level of abstraction, \sol{} works as follows. Each ground
station runs a small bank of capacity-diverse learning models, and picks one per event from the channel condition that delivered it. The selected model's
score becomes a flag through a streaming statistical fit at the
station, so the threshold tracks the drift in the score
distribution. Across stations, only classical parameters move, weighted by how reliable each station's
hardware is, so noisy backends do not drag the global
model. Events arriving on a busy QPU are dropped, and offline
stations do not contribute.

\vspace{-2mm}
\subsection{Design Challenges and Requirements}
\label{sec:design:reqs}

\sol{}'s goal is to learn on quantum-sensed data without any
classical intermediate, from the orbital probe through to the ground
inference. Training and deployment share the same channel,
QPU mix, and drop policy, so the system operates within a
single regime that subsumes both. \add{This sharing is forced --  a sensed state can be neither stored nor labeled, so no offline training set can exist, and learning has to happen online, on the stream the system serves.} Five challenges shape this regime,
none of which \add{are solved by classical remedies}, shaping \sol{}'s
requirements.

\vspace{2mm}

\noindent\textbf{C1. Teleported states cannot wait.}
Qubit coherence times are orders of magnitude below the gaps between
contact windows (Sec.~\ref{sec:bg:coordination}), and the no-cloning
theorem~\cite{wootters2009no} of quantum states forbids buffering. A sensed state that arrives
while the receiving QPU is occupied is lost \add{since there is no queue}. Stations can also go offline due to weather or maintenance.

\vspace{2mm}

\noindent\textbf{C2. The sensor itself is noisy.}
Platform-specific imperfections (Sec.~\ref{sec:bg:sensing}) corrupt the
state before it enters the link, and the noise structure differs
across sensor modalities. The input the learner receives is already mixed at the source.

\vspace{2mm}

\noindent\textbf{C3. The atmospheric channel varies per event.}
The Pauli error rates each downlinked qubit incurs depend on
satellite position, slant geometry, and the station's local
turbulence profile. A single station
encounters a different effective channel from one event to the next,
and different stations observe different distributions over their
respective passes. No single model capacity remains adequate across
this range.

\vspace{2mm}

\noindent\textbf{C4. Ground hardware is heterogeneous and noisy.}
Each station hosts a physically distinct QPU with its own gate
errors, coherence times, and coupling map, and these characteristics
drift between recalibration windows. A uniform treatment of every
station's gradient contribution can let the noisiest backends bias
the global model.

\vspace{2mm}

\noindent\textbf{C5. Labels are not reliably available.}
Labeling a sensed state at the satellite requires destructive
measurement, and no-cloning forbids retaining a copy for downstream
use.

\vspace{1mm}

The challenges interact. First, C1, C2, and C3 jointly rule out approaches that increase per-event compute to compensate for channel noise. Under the no-buffering constraint, additional compute per event increases the drop rate over the contact window.
The capacity dimension compounds this. A model sized for mild
channels overfits the broadening from heavy dephasing and
depolarization, and a model robust to heavy noise underfits clean
channels.
Second, \add{C2--C4} introduce continuous drift. Sensor, channel, and backend noise vary across events, which a fixed rule fails to track.

\add{C1 and C5 have no classical analogs. For C2--C4, classical systems calibrate or denoise sensors, retransmit after channel errors, and reweight stored contributions from heterogeneous hardware. All of these need data that can be stored, copied, or re-sent, which a sensed state cannot, so noise should be handled on the state itself, within the per-event budget (Sec.~\ref{sec:design:system})}.
These challenges and their interactions impose six \textit{requirements} on
\sol{}.
C1 forbids buffering, so the pipeline has to tolerate dropped
events \textbf{(R1)}.
C5 forbids per-event labels, so the system has to operate without
external supervision \textbf{(R2)}.
The cross-station heterogeneity and drift of C4 make uniform
aggregation unsafe, so contributions need a per-station trust
signal refreshed each round \textbf{(R3)}.
C1, C2, and C3 jointly leave no compute headroom per event for
noise compensation, so the system has to maintain robustness to
input and channel noise within a fixed per-event budget
\textbf{(R4)}.
Continuous drift across C2--C4 leaves any fixed reference stale,
so the system has to track its operating distribution online
\textbf{(R5)}.
Stations can also go offline through weather, maintenance, or
recalibration (C1, C4), so the system has to withstand station
downtime and backend recalibration without compromising the global
model \textbf{(R6)}.


\begin{figure}[t]
  \centering
  \includegraphics[width=0.99\columnwidth]{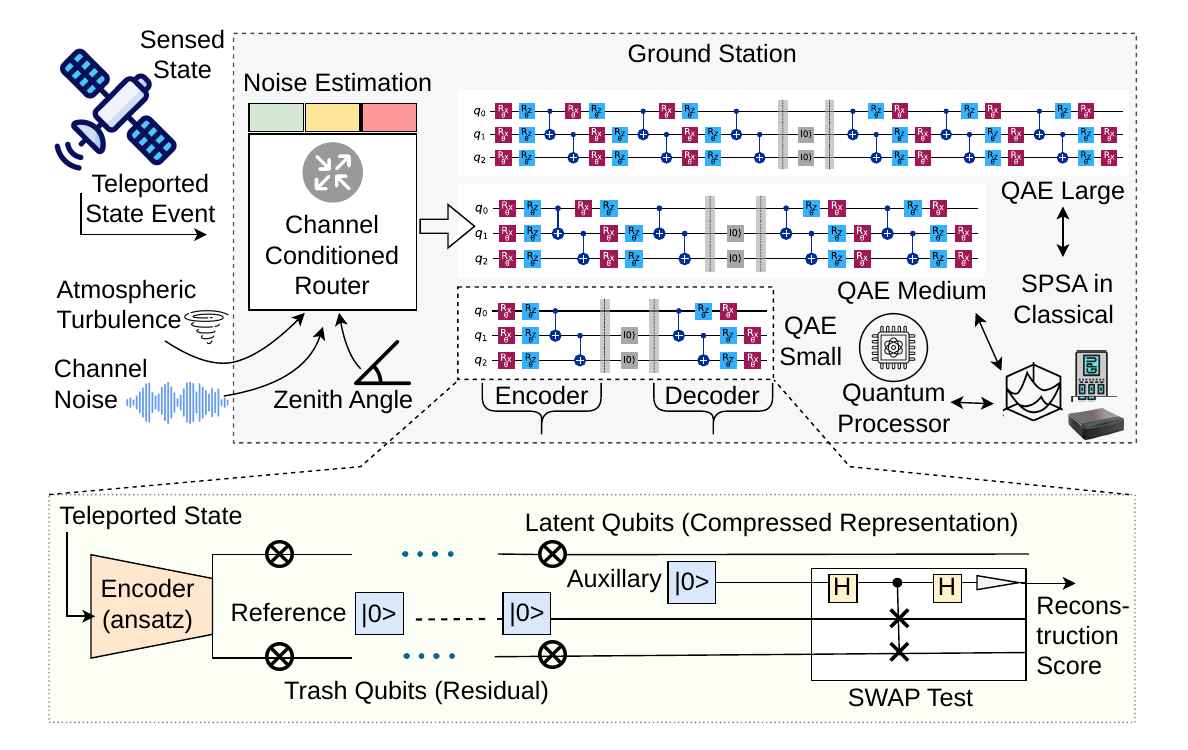}
  \vspace{-4mm}
  \caption{\textit{Classical telemetry alone selects one capacity-tier QAE on the QPU, with SPSA gradient estimation on a classical node. The inset details the label-free SWAP-test reconstruction, which yields a scalar score without measuring the input state.}}
  \vspace{-4mm}
  \label{fig:qae}
\end{figure}

\vspace{-2mm}
\subsection{The \sol{} Distributed System}
\label{sec:design:system}

The primary contribution of \sol{} is \add{a distributed system that learns from sensed quantum data that cannot be stored, copied, or re-sent, and that arrives over an intermittent noisy link.} It targets orbital quantum
sensing workloads of the kind introduced in Sec.~\ref{sec:bg}. These workloads share
three properties. Streams are mostly nominal with a minority of anomalous events. Labels at the source require destructive measurement
and are not available. The atmospheric channel and the ground
hardware vary continuously across events and across ground stations.

\sol{} addresses these with two components. Each ground station runs a
per-event pipeline. A federated coordination layer maintains a
single global model across stations by sharing only classical
model parameters between rounds. Two design decisions distinguish
\sol{} from a basic federated quantum learning setup. At the
station, a capacity-diverse model bank with channel-conditioned
per-event routing handles a channel range that no single model
capacity covers. Across stations, a quantum fidelity-weighted circular
aggregation handles heterogeneous and drifting QPUs that a uniform
federated average cannot. 



\subsubsection{The Per-Event Pipeline at a Ground Station}
\label{sec:design:perevent}
\label{sec:design:unsup}
\label{sec:design:qaebank}

At each station, every event undergoes the same processing (Fig.~\ref{fig:qae}).
Station telemetry produces a channel-condition estimate. The
estimate selects one of several models held at the station. The
selected model produces a reconstruction score from the
teleported quantum state. A streaming threshold converts the
score into a binary flag.

\vspace{1.5mm} 

\noindent\textbf{Label-free reconstruction via a Quantum Autoencoder.}
The workloads with quantum orbital sensing use cases fall in the category of unsupervised anomaly detection problems for
two reasons. First, no labels are produced at the source.
Labeling at the satellite requires destructive measurement (C5),
which is incompatible with transmitting the quantum state for
downstream processing. Simulated anomaly labels at training
time diverge from what the orbital sensor produces in
deployment, and a model trained on them is miscalibrated against
the real stream. Second, the channel, the sensor, and the ground
hardware drift continuously, so the system has to keep learning
over its lifetime and keep segregating normals from anomalies as
the score distribution shifts under that drift. \sol{}
operates on the unlabeled stream the system observes (R2), and
the training signal has to come from a quantity the model itself
produces on each event.

Among unsupervised approaches, three standard scalar signals
exist for anomaly detection. Reconstruction error from an
autoencoder asks whether each input can be compressed through a
bottleneck and recovered. Likelihood under a generative model
asks whether each input is typical under a learned density.
Distance from a learned centroid asks whether each input sits
near a learned reference state. \sol{} uses reconstruction
error. Of the three, it is the only one that imposes a
structural compression constraint on the model. Each input has
to be recoverable from a compact latent representation, which
gives the model an inductive bias toward the structure of the
normal distribution. Likelihood estimation requires an explicit
density model that is expensive to learn on quantum states. A
learned centroid requires a metric on states and a single
anchor, which fits a normal distribution only at low spread and
collapses under the broader normal distribution this workload
produces. The per-event scalar from reconstruction at training
is also the scalar the threshold consumes at deployment.

For quantum-sensed data, the autoencoder has to be a Quantum
Autoencoder (QAE)~\cite{romero2017quantum}. A classical
autoencoder would receive a classical feature vector from the
satellite, which requires measuring the quantum state at the
source and collapses the entanglement structure that separates
normal from anomalous states (Fig.~\ref{fig:motivation_f1}).
The QAE applies a parameterized encoder to the quantum state (Fig.~\ref{fig:qae}).
The encoder partitions the result into latent qubits (compressed representation) and trash qubits (residual).
When compression succeeds, the trash qubits carry no
information about the input and approach a fixed reference
state. Reconstruction quality is defined as the closeness of the
trash qubits relative to that reference state.

\begin{figure}[t]
  \centering
  \includegraphics[width=0.99\columnwidth]{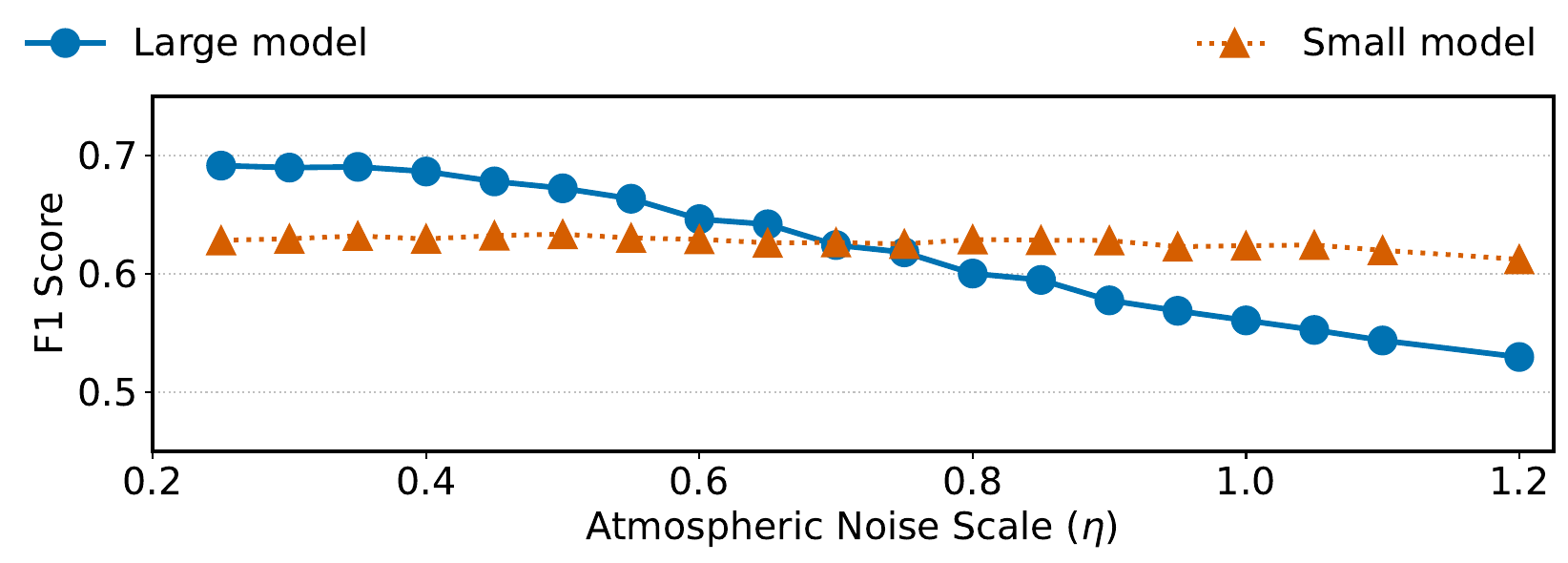}
  \vspace{-4mm}
  \caption{\textit{F1 score against the per-link channel noise
  parameter $\eta$ for two QAE capacities at opposite ends of
  the model-size axis. The large capacity has higher F1 at low
  $\eta$. The small capacity has a higher F1 at high $\eta$. No
  single capacity covers the range. The pattern generalizes to other pairs on the axis. }}
  \vspace{-6mm}
  \label{fig:design-crossover}
\end{figure}

Reading out the trash qubits directly would destroy the quantum
state and prevent downstream use of the latent representation.
The QAE instead uses a SWAP test, a quantum primitive
that returns the closeness between two quantum registers as a
single scalar without disturbing the registers it compares. The
test uses an auxiliary qubit and a controlled-SWAP between the
trash register and the reference register. A single measurement
on the auxiliary qubit produces an outcome whose probability grows as the trash register deviates from the reference.
This rate is the per-event reconstruction score, low when the encoder has compressed the input into the latent register and high otherwise. No projective measurement of the input itself occurs, so the
entanglement structure of the teleported state is preserved end to end.

\vspace{1.5mm}

\noindent\textbf{Capacity-diverse multi-model bank with channel conditioned routing.}
R4 requires robustness across the per-event channel range within
a fixed compute budget, and no single QAE capacity provides it.
Fig.~\ref{fig:design-crossover} shows this for two QAE
capacities at opposite ends of the model-size axis. F1 score is
plotted against the per-link channel noise parameter $\eta$.
The large capacity has a higher F1 at low $\eta$, where the model
has headroom to learn the reconstruction structure, and degrades
at high $\eta$, where it overfits the broadening that heavy
dephasing and depolarization induce. The small capacity is flat
across the range and outperforms the large capacity at high
$\eta$. The pattern generalizes to any pair across the
model-size axis, and our evaluation confirms this.

\sol{} therefore maintains a capacity-diverse bank of QAEs at
each station. The router selects the bank member from a channel-condition estimate,
\vspace{1mm}
\begin{equation}
\hat{N}_{\text{ch}}(\zeta, A_s, \eta) \;=\;
   \tfrac{\eta}{2}\bigl(1 - e^{-\sigma_R^2(\zeta, A_s)}\bigr),
\label{eq:nch}
\end{equation}
\vspace{1mm}
The equation combines two factors. The first, $\eta/2$, is the
per-link noise scaling (the $\eta$ swept in
Fig.~\ref{fig:design-crossover}). The second,
$(1 - e^{-\sigma_R^2(\zeta, A_s)})$, is the atmospheric
scintillation factor. $\sigma_R^2$ here is the Rytov variance,
a standard measure of optical turbulence along the line of
sight. It is computed from the station-satellite zenith angle
$\zeta$ and the station's atmospheric turbulence strength $A_s$
through the Hufnagel--Valley refractive-index profile (discussed
in Sec.~\ref{sec:methodology}). Clean links give
$\hat{N}_{\text{ch}} \to 0$. All three inputs are obtained
without invoking the quantum state, so the router commits to a
model before any quantum operation runs. The
$\hat{N}_{\text{ch}}$ range is partitioned into zones
bijectively mapped to the bank members. Low
$\hat{N}_{\text{ch}}$ routes to a higher-capacity member and
high $\hat{N}_{\text{ch}}$ to a lower-capacity member. This
zone-to-member mapping follows the capacity-noise pattern of
Fig.~\ref{fig:design-crossover}. The bank's members are
trained on the same federated distribution under the same
atmospheric, sensor, and QPU noise. Capacity is the only
variable that differs across them, so each is a generalist at
its scale and not a specialist for a single channel band. The selected model's score is the flag for the event, but the
other bank members still run. Each scores the event, takes a
local optimizer step against the reconstruction loss, and
updates its own streaming threshold. Without this, an idle
model's weights and threshold would drift away from the channel
statistics, and when conditions later brought it back into
routing range, neither would match the live distribution. The
compute cost of keeping all members active is reserved in the
per-event budget. The router commits to one model and does not
blend predictions across the bank (Fig.~\ref{fig:qae}). Different channel conditions
are best served by different capacities, and a blended ensemble
does not recover the capacity advantage at either end. Hard
routing also avoids the extra hyperparameters a soft-mixture or
overlapping-zone scheme would introduce.

\begin{figure}[t]
  \centering
  \includegraphics[width=0.99\columnwidth]{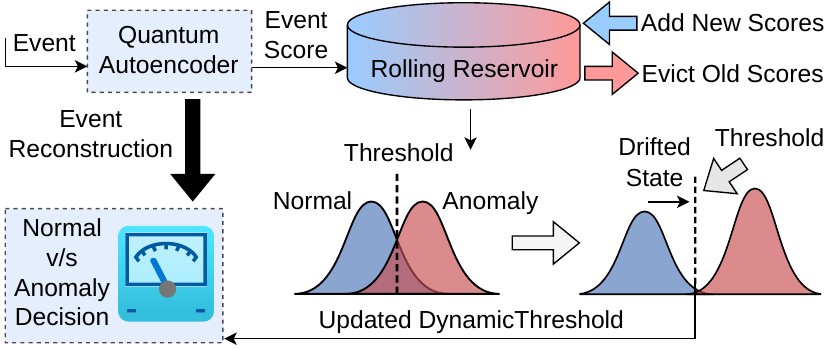}
  \vspace{-4mm}
  \caption{\textit{Per-station QAE supplies both the score and, via the GMM-2 refit on the rolling reservoir, the threshold the score is judged against, so classification adapts to drift without labels.}}
  \vspace{-4mm}
  \label{fig:dynamic_threshold}
\end{figure}

%


\vspace{2mm}

\noindent\textbf{Label-free dynamic thresholding.}
A trained bank member produces low reconstruction scores for
samples it can reconstruct (normals) and high scores for samples
it cannot (anomalies), so the score distribution has two
clusters. The bank is trained on a normal-dominant mix without
filtering anomalous signal, so both clusters have non-zero
spread and overlap near the boundary. A fixed cutoff fails
because channel and QPU drift shift both cluster means and
spreads. Three timescales of drift apply. The atmospheric
channel varies per event with the satellite's geometry, the
local turbulence, and the weather.
The orbital sensor's own noise drifts more slowly. The ground
QPU drifts between recalibration windows. The two-cluster structure motivates a two-component mix
model for score distribution, one component per cluster.
\sol{} uses a Gaussian for each component, since each cluster
is unimodal and Gaussian parameters fit from observed scores in
closed form. The resulting 2-component Gaussian mixture
(GMM-2) places the threshold $\tau$ at the score where the two
component densities cross,
\begin{equation}
w_L\,\mathcal{N}(\tau; \mu_L, \sigma_L^2) \;=\;
w_H\,\mathcal{N}(\tau; \mu_H, \sigma_H^2),
\end{equation}
where $\mathcal{N}(\tau; \mu, \sigma^2)$ is the Gaussian density
at $\tau$ with mean $\mu$ and variance $\sigma^2$, $w_L$ and
$w_H$ are the mixing weights of the low-score (normal) and
high-score (anomalous) clusters, $\mu_L, \mu_H$ are their
cluster means, and $\sigma_L^2, \sigma_H^2$ are their cluster
variances. The crossing is the score at which a sample is
equally likely under either component, and it has a closed-form
solution, a quadratic in $\tau$ with the relevant root in
$(\mu_L, \mu_H)$. Both components are re-estimated from
observed scores. The crossing moves
with them and the threshold tracks drift (Fig.~\ref{fig:dynamic_threshold}). Each station maintains a rolling reservoir of recent
reconstruction scores, indexed per station because each
station's geometry produces a distinct distribution per
model, as the bank members score differently. The GMM-2 is
re-fit periodically from this reservoir, not on every event.
Per-event refit would track short bursts of noisy events and
oscillate. New events evict
oldest scores, refreshing the fit.

\vspace{-2mm}
\subsubsection{Federated Coordination Across Stations}
\label{sec:design:federation}
\label{sec:design:esp}

The per-event pipeline runs independently at every station, but
a single model that generalizes across the orbit has to
incorporate data from all stations. Different stations receive
events from different satellite passes (Fig.~\ref{fig:coverage}),
and relaying sensed states to a central station is not viable.
A second quantum hop compounds the atmospheric noise, and
the satellite cannot measure the state for classical
transmission without collapsing the entanglement structure
QAE requires. \sol{} therefore trains
in parallel at each station and exchanges only classical model
parameters between rounds.

Each training round begins with the server broadcasting the
current global weights $\theta^{(t)}_{\text{global}}$ to all
stations. The station scores its incoming events through its model bank, takes local optimizer steps to
reduce the SWAP-test costs, and at round close returns its
updated weights together with its ESP signal (defined below).
The server combines the contributions via ESP-weighted FedAvg,
and the cycle repeats.

\vspace{1.5mm}

\vspace{1mm}
\noindent\textbf{ESP-weighted circular aggregation.}
Noisier QPUs produce contaminated gradient
updates, and uniform FedAvg that treats all stations equally
would let those updates bias the global model. The aggregator
therefore needs a per-station trust signal. The signal has to
be computable without labels, refresh per round to track
recalibration (R6), and capture gradient quality instead of
sample count.

\begin{figure}[t]
  \centering
  \includegraphics[width=0.99\columnwidth]{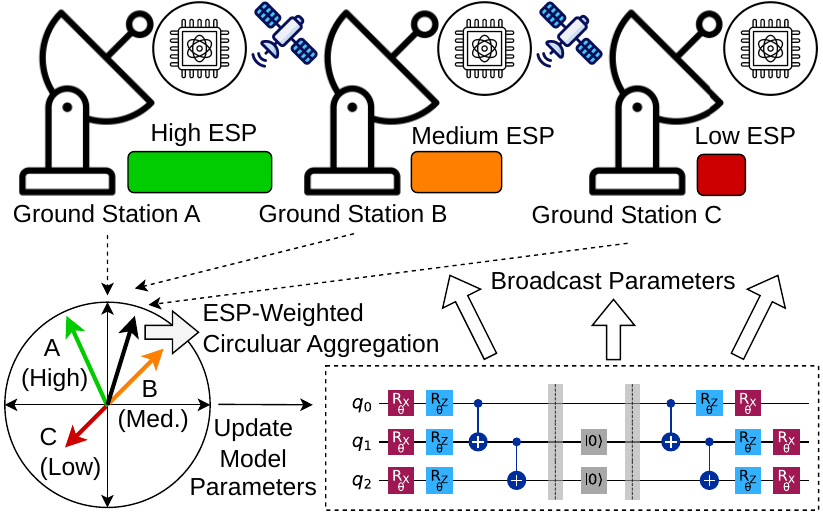}
  \vspace{-4mm}
  \caption{\textit{Stations return gate-angle parameters weighted by their expected success probability (ESP). The global update is their ESP-weighted circular mean (the vector sum on the circle), so a low-ESP station pulls it less.}}
  \vspace{-5mm}
  \label{fig:circular}
\end{figure}

We use Expected Success Probability (ESP), the
calibration-derived probability that a circuit executes
correctly on a given backend. ESP is the product of three terms
covering gate errors, amplitude relaxation, and dephasing, defined as -- 
$
\text{ESP} \;=\; \prod_{i \in \text{gates}}(1 - \epsilon_i)
            \cdot e^{-d\,t_g / T_1}
            \cdot e^{-d\,t_g / T_2},
$ 
where $\epsilon_i$ is the calibrated error rate of gate $i$, $d$
is circuit depth, and $t_g$ the average gate duration. ESP is
computed from each backend's live calibration and is cheap to
recompute per round. Alternatives like total variation
distance (TVD) need the ideal output distribution per circuit, far more expensive to characterize. At round $t$, the aggregator combines each station $k$'s
locally trained weights $\theta_k^{(t)}$ with mixing weight
proportional to that station's ESP. The trainable
parameters are gate rotation angles, so aggregation is on the
unit circle,
\begin{equation}
\begin{aligned}
&\theta_{\text{global}}^{(t+1)} \;=\;
  \mathrm{atan2}\!\left(
    \textstyle\sum_k w_k(t) \sin\theta_k^{(t)},\;
    \sum_k w_k(t) \cos\theta_k^{(t)}\right), \\
&\qquad w_k(t) \;=\; \frac{\text{ESP}_k(t)}{\sum_j \text{ESP}_j(t)}.
\end{aligned}
\end{equation}
An arithmetic mean breaks down for angles near the $\pm\pi$
wraparound. The circular form is topologically correct (Fig.~\ref{fig:circular}). ESPs
are recomputed per round against each station's live
calibration, so the mixing weights track QPU drifts.

\vspace{1.5mm}

\noindent\textbf{Robustness to station outages and QPU recalibration.}
Stations can go offline through weather, maintenance, or
hardware faults. \sol{} handles this in two ways. Events the
offline station would have received are rerouted to the nearest
operational station whose coverage window includes the pass; otherwise they are dropped (R1).
The federated round proceeds with the reduced station set, and
ESP-weighted aggregation renormalizes over the remaining
contributors. When the station returns, it pulls the current
global parameters in the next round and reinitializes its
streaming threshold from incoming scores.
QPUs can recalibrate between events. An IBM backend, for
example, may expose a new coupling map and updated gate errors.
\sol{} tolerates recalibration because the trained QAE
parameters are the rotation angles. On a recalibration, the station retranspiles the
circuit against the live backend and continues with the same
parameters. The ESP signal incorporates the new gate errors
and propagates them into the next round's aggregation weights
without retraining.

\vspace{-1mm}
\subsubsection{Scope and Generality}
\label{sec:design:scope} 
\add{
Teleporting sensed states to multiple ground stations is tied to the satellite setting, as no single station sees the whole orbit (Fig.~\ref{fig:coverage}). The model bank applies to any noisy quantum link, terrestrial or orbital. One model capacity cannot serve all events (Fig.~\ref{fig:design-crossover}). Fidelity-aware aggregation applies to any heterogeneous QPUs that share one model. These three mechanisms are not an arbitrary composition. Each one is forced by a constraint. None of these mechanisms reads a task's labels or metric. Hence, the pipeline supports the sensing tasks of Sec.~\ref{sec:intro} beyond the anomaly detection and classification we evaluate (Sec.~\ref{sec:method:dataset}). If the QPU were placed in orbit, only the satellite-to-ground link would go away. Federation would run across satellites, the model bank would be conditioned on sensor noise, and fidelity-weighted aggregation would stay because QPUs drift.
}

\vspace{-2mm}
\subsection{Putting It All Together}
\label{sec:design:together}
End-to-end, the orbital sensor prepares its qubits in the sensing
superposition, exposes them to the field, and teleports each
phase-encoded state to the ground station whose contact window covers
it (Sec.~\ref{sec:bg:downlink}), where the station applies the
radio-link Pauli correction to recover it on its QPU. The QAE is a
parameterized ansatz (Sec.~\ref{sec:bg:qc}) whose rotation-gate angles
$\boldsymbol{\theta}$ are its trainable parameters. Before any quantum
operation, the channel-conditioned router selects the bank member for
the event (R4). A new event is dropped if the QPU is busy (R1), since
coherence times forbid buffering. Otherwise the selected QAE produces
the SWAP-test reconstruction score (R2) and the streaming GMM-2
threshold converts it to a binary flag (R5). Everything else runs on a
classical compute node co-located with each station, which updates
$\boldsymbol{\theta}$ from the per-event reconstruction loss via the
SPSA gradient estimation (\S\ref{sec:method:training}), maintains the rolling
score reservoir, performs the periodic GMM-2 refit, and exchanges
parameter vectors with the federation server. At round close, every
station returns its updated $\boldsymbol{\theta}$ and its ESP from live
calibration. ESP-weighted circular aggregation merges the
contributions on the unit circle (R3), and the merged global model
broadcasts at the next round's start.

\sol{} is an online learning system that runs training and inference
jointly on every event. After the selected QAE scores an event, its
local optimizer takes a training step against the reconstruction loss,
and the other bank members also score it and take their own training
steps so they stay calibrated to the live channel. Each score enters
the rolling reservoir that feeds the periodic GMM-2 refit, so the
threshold at event $t+1$ reflects scores through event $t$ and the
model has taken one more step than at event $t$. \sol{} thus tracks the three
drift timescales the design assumes -- the per-event channel, the slower
orbital sensor, and the inter-recalibration QPU. Under station outages
and QPU recalibrations (R6), failover reroutes affected events with ESP
renormalization, and recalibration retranspiles the ansatz against the
live backend without retraining.

\vspace{-2mm}
\section{Implementation and Methodology} 
\label{sec:methodology}

\subsection{Hardware setup}
\add{
\sol{} combines real hardware with simulation. The orbits come from published orbital data, and the orbit-to-ground link is simulated. Table~\ref{tab:realvssim} summarizes our prototype.}

\vspace{-2mm}
\begin{table}[!htb]
  \centering
  \caption{\textit{Real and simulated parts of our hybrid prototype.}}
  \vspace{-4mm}
  \label{tab:realvssim}
  \scalebox{0.75}{
  \begin{tabular}{@{}>{\raggedright\arraybackslash}p{0.25\columnwidth}>{\raggedright\arraybackslash}p{0.48\columnwidth}>{\raggedright\arraybackslash}p{0.48\columnwidth}@{}}
    \toprule
    \rowcolor{headergray}
    \textbf{Stage} & \textbf{Real hardware} & \textbf{Simulated} \\
    \midrule
    Sensor input & Aquila noise channel, measured on device (Sec.~\ref{sec:method:sensor}) & Sensed state $\rho$, encoded from telemetry (Sec.~\ref{sec:method:dataset}) \\
    \midrule
    Ground-station QPUs & IBM Heron and Nighthawk (Sec.~\ref{sec:method:gs}) & QPU-noise scaling sweep only (Sec.~\ref{sec:eval:robustness}) \\
    \midrule
    Orbit-to-ground link & --- & Entire downlink (Sec.~\ref{sec:method:channel}) \\
    \bottomrule
  \end{tabular}
  }
  \vspace{-3mm}
\end{table}

\vspace{-2mm}
\subsubsection{Satellite and ground-station QPUs}
\label{sec:method:gs}
The configuration spans up to eight low-Earth-orbit satellites carrying
quantum sensors and up to eight ground-station QPUs.
Orbits are propagated from published two-line element sets with the standard SGP4 propagator \cite{vallado2006revisiting}.

The main experiments use four stations, Fairbanks, Wallops, Madrid, and Singapore,
which run on real superconducting IBM hardware (Heron and Nighthawk).

\vspace{-2mm}
\subsubsection{Satellite-to-ground quantum channel}
\label{sec:method:channel}
The transmitted state crosses the atmosphere
to the receiving QPU and arrives with turbulence-induced channel noise \cite{trinh2022statistical}.
We model it as a per-qubit composition of two completely positive trace-preserving (CPTP)
maps, with strengths fixed per event by slant-path turbulence, which follows the Hufnagel--Valley (HV) profile
\cite{valley1980isoplanatic, andrews2005laser}, set by a per-station
boundary-layer strength $A_s$ (m$^{-2/3}$) and standard in satellite
quantum-channel work \cite{bourgoin2013comprehensive, ghalaii2022quantum,
behera2024estimating, gonzalez2024satellite}. From it we compute the Rytov
variance $\sigma_R^2(\zeta, A_s)$ through the slant-path integral
\cite{andrews2005laser}, with $\zeta$ taken from SGP4
\cite{vallado2006revisiting}, and convert it
via the exponential decay of coherence with accumulated phase variance \cite{andrews2005laser}, to per-qubit dephasing ($q$) and depolarization ($p$) channels applied to every qubit of the $n$-qubit density matrix $\rho$ \cite{nielsen2010quantum}.
\add{This link is simulated in our prototype for evaluation.}

\vspace{-1mm}
\subsubsection{Input sensor}
\label{sec:method:sensor}
For the main experiments, we assume the density matrix $\rho$ as the sensing input,
since no orbital quantum sensor is yet deployed. 
\vspace{1mm}

\begin{figure}[t]
  \centering
  \includegraphics[width=0.95\columnwidth]{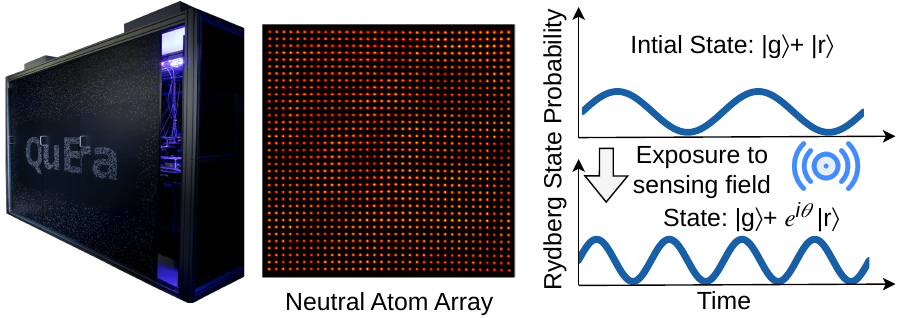}
  \vspace{-4mm}
  \caption{\textit{Field sensing on QuEra Aquila. Atoms start in a ground--Rydberg superposition $|g\rangle+|r\rangle$, and the field imparts a
phase of $\theta$ onto $|r\rangle$ that encodes the reading.}}
  \vspace{-5mm}
  \label{fig:sensor-cartoon}
\end{figure}

\noindent\textbf{Neutral atom sensor.}
We use QuEra Aquila \cite{wurtz2023aquila}, a real neutral atom
device whose Rydberg states respond sharply to weak fields and stay coherent
long enough to read the shift out (Sec.~\ref{sec:bg:sensing}). A controlled
Hamiltonian prepares each atom in a ground--Rydberg superposition. The sensed field shifts the atom's transition frequency, so
over the exposure the relative phase winds at the
shifted rate while the populations stay fixed. The accumulated phase, read out
by interference, is the field value, and longer exposure within the
coherence time gives a finer reading (Fig.~\ref{fig:sensor-cartoon}). From Aquila's calibration we fit a three-component noise spectral density
\cite{ludmir2026spintune} to Ramsey and CPMG sequences on \add{the Aquila device through AWS Braket}, and derive a single-atom
Kraus channel, amplitude damping at its spontaneous-emission $T_1$ and
dephasing matched to the coherence loss over the $\tau = 4\,\mu\text{s}$
exposure window. We apply it to each of $\rho$'s atoms before the atmospheric channel (Sec.~\ref{sec:method:channel}), so an event accumulates sensor, channel, and QPU noise in turn.

\subsection{Model and training}

\subsubsection{Model bank}
\label{sec:method:bank}
The capacity-diverse bank (Sec.~\ref{sec:design:qaebank}) holds three quantum
autoencoders, large ($L$), medium ($M$), and small ($S$). All use an \texttt{EfficientSU2} \cite{ibmEfficientSU2latest} encoder with
full entanglement and alternating $R_y$/$R_z$ rotations, followed by the SWAP
test of Sec.~\ref{sec:design:qaebank}. $L$ uses four qubits at depth four, $M$ three
qubits at depth three, and $S$ three qubits at depth one. $L$ carries a
two-qubit trash space compared against $\ket{0}^{\otimes 2}$, and $M$ and $S$ carry a
single trash qubit against $\ket{0}$.

\vspace{-1mm}
\subsubsection{Training}
\label{sec:method:training}
Each client trains its local copy with Adam at
learning rate $0.03$ on the IBM hardware (Sec.~\ref{sec:method:gs}), at $1024$ shots. Since the parameter-shift rule of
Sec.~\ref{sec:bg:qc} is
prohibitively expensive. We instead estimate the gradient with
Simultaneous Perturbation Stochastic Approximation (SPSA)
\cite{spall2002multivariate}, a standard estimator under hardware noise \cite{kandala2017hardware, wiedmann2023empirical,
filip2024fighting}. We track the area under the precision-recall curve
(PR-AUC) across rounds, a threshold-independent metric for anomaly detection
\cite{davis2006relationship, saito2015precision, 10.14778/3538598.3538602}.

\vspace{-1mm}
\subsection{Datasets and encoding}
\label{sec:method:dataset}
We evaluate on OPS-SAT \cite{ruszczak2025ops} of Sec.~\ref{sec:bg} and SMAP
\cite{10.1145/3219819.3219845}, widely used satellite telemetry benchmarks. Each
segment is summarized by a handful of statistical features and fit to the
$2^n$-length amplitude vector an $n$-qubit member expects.

OPS-SAT ships as pre-extracted segments, while SMAP is raw per-channel
telemetry. For SMAP, we take a subset of channels and cut each into
non-overlapping windows, summarized with the same features as OPS-SAT \cite{ruszczak2025ops}. Since SMAP anomalies are defined
per channel, we standardize features per channel against that channel's
anomaly-free reference and draw a split at OPS-SAT's anomaly rate ($\approx$20\%). The same channels, windows, features, and split serve
\sol{} and every baseline. The one SMAP-specific change is the encoding: it keeps each window's magnitude.

\vspace{1mm}

\noindent\textbf{Supervised image tasks.}
We also run classification tasks, MNIST (3 versus 4) and FashionMNIST (bag versus
sandal), for a generality check. The federated topology, noise, and router banks are unchanged; only the SWAP-test reconstruction objective becomes a supervised readout. Each $28\times28$ image is flattened as above, into a balanced $\sim$1600/400 split. \add{The other sensing tasks named in Sec.~\ref{sec:intro} are supported but not evaluated (Sec.~\ref{sec:design:scope}).}

\vspace{-2mm}
\subsection{Competing Solutions}


\noindent\textbf{Isolated.}
Each station runs its own QAE on its own shard under its own channel and QPU noise
and shares nothing.


\noindent\textbf{Centralized.}
Every station forwards its sensed states to one server, which trains a single
QAE on the pooled data under one QPU's calibration. Each state crosses the
satellite-to-ground link and a second, lossier ground-to-central hop, so the
central model trains on doubly attenuated states.

\noindent\textbf{FedAverage.}
The federated protocol with plain FedAvg, replacing the ESP-weighted
aggregation of Sec.~\ref{sec:design:esp}.

\vspace{1mm}

\noindent\textbf{Quorum~\cite{ludmir2025quorum}.}
\label{sec:method:quorum}
Quorum is a standalone quantum autoencoder anomaly detector with no notion of
an orbital channel, sensing, or federation, but it is the closest prior quantum
method, so we port it into \sol{}'s setting. It scores through
SWAP-test reconstruction but trains no encoder: it fixes an ensemble of
encoders and scores each sample by its reconstruction-cost deviation from a bucketed ensemble mean. We run it
on the same channel and per-station QPU noise as \sol{}, with a
larger shot budget.

Two of its mechanisms cannot run on a single teleported state. Its score needs
a second copy of the input to SWAP-test against, which no-cloning forbids
\cite{wootters2009no}, so we use \sol{}'s single-state trash-qubit SWAP test
(Sec.~\ref{sec:design:qaebank}). It also re-selects a random feature subset per
member, but the ground station holds only the sensed state, so we drop feature
selection. We otherwise follow Quorum as published and bucket its events by the
same $\hat{N}_{\text{ch}}$ the router uses (Sec.~\ref{sec:design:qaebank}).

\subsection{Evaluation Protocol}
\label{sec:method:eval}
\vspace{1mm}
\noindent\textbf{Event stream.} 
We evaluate on a real-telemetry event stream over one orbital window across
the constellation of satellites and ground stations
(Sec.~\ref{sec:method:gs}). We encode the data (Sec.~\ref{sec:method:dataset})
in the neutral atom quantum sensors \add{as the sensors' output state $\rho$} and treat it as the sensing input (Sec.~\ref{sec:method:sensor}). We then simulate the orbital
sensor-to-ground-station pass, propagating the orbit and applying the
atmospheric channel (Sec.~\ref{sec:method:channel}) to each downlinked state.
All events are drawn from the held-out test partition, and every method is
scored on the same stream.

\vspace{1mm}

\noindent\textbf{Serving model.}
We model each station as an M/D/1/1 server, so any event that
arrives while it is busy is dropped\add{, the drop policy of C1 (Sec.~\ref{sec:design:reqs})}. We use IBM superconducting QPUs to process the events. An event's service time is one inference
plus online training over the models the method runs on.

\vspace{1mm}

\noindent\textbf{Metrics.}
Our system metrics are throughput, latency, and processor utilization, with
utilization normalized to the in-view active window, and we also track drop rate. For detection quality, we report F1 score,
precision, recall, and accuracy, with F1 as the primary measure. \add{F1 is the metric of the anomaly-detection task (not of \sol{}'s mechanism (Sec.~\ref{sec:design:scope})).} Isolated is
the baseline, so we report both the percent improvement over it and the raw metric values.



\vspace{3mm}
\section{Evaluation}
\label{sec:eval}

We first establish the end-to-end effectiveness of \sol{}, then examine its robustness, scaling, and the ablations that explain why it works in practice.

\vspace{-2mm}
\subsection{Effectiveness of \sol{}}

\begin{figure}[t]
  \centering
  \includegraphics[width=\columnwidth]{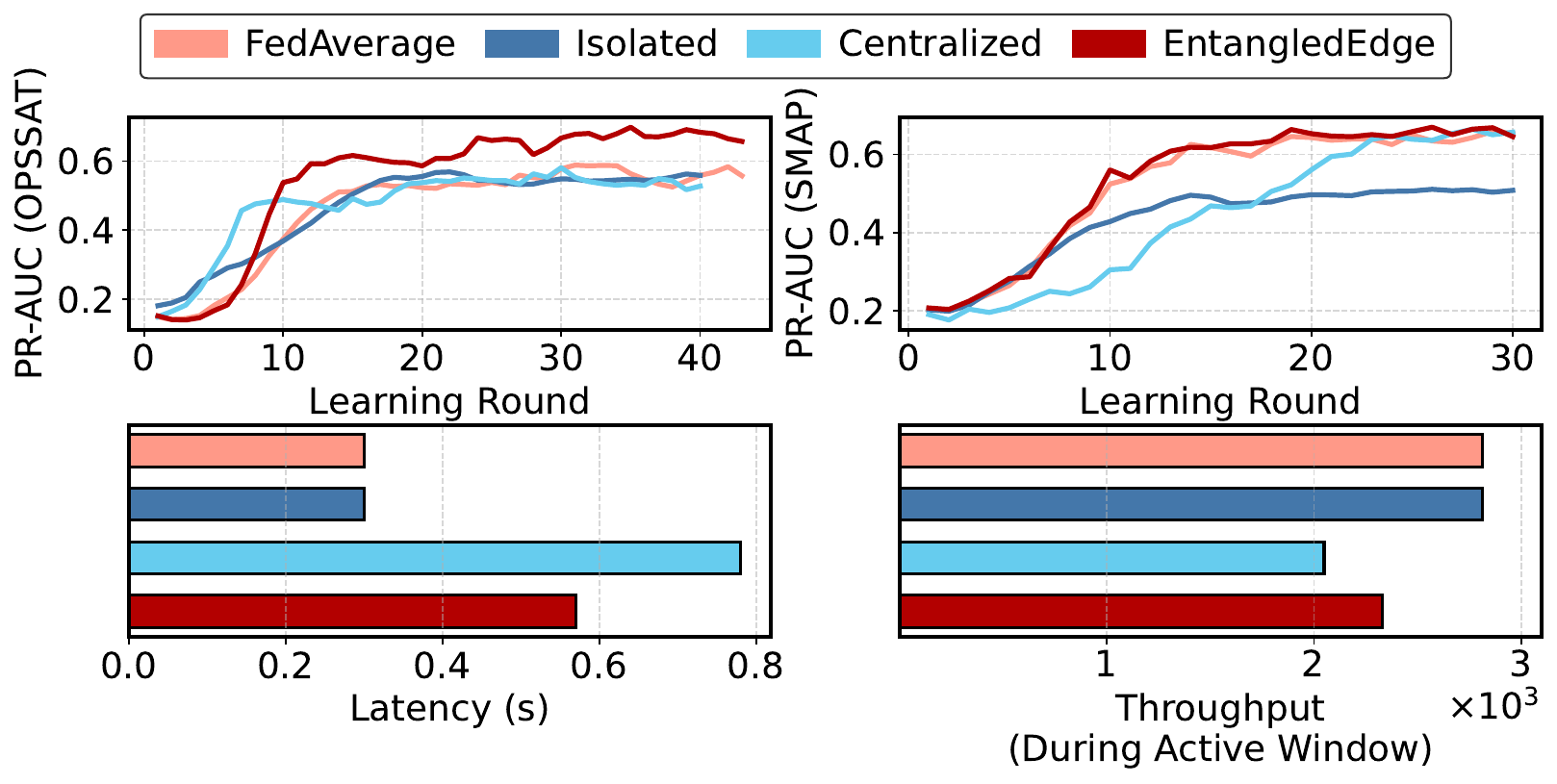}
  \vspace{-8mm}
\caption{\textit{\sol{} converges to the top PR-AUC rank on both datasets (top), and serves events at latency and throughput near the single-model baselines, while centralized is slowed by its teleportation hop (bottom).}}
  \label{fig:flagship_system}
  \vspace{-2mm}
\end{figure}

We compare \sol{} against the isolated, centralized, and FedAverage baselines on the
held-out stream, scored in streaming mode (Sec.~\ref{sec:method:eval}). All four
converge in PR-AUC on both OPS-SAT and SMAP (Fig.~\ref{fig:flagship_system},
top), and \sol{} reaches the top rank on each, with FedAverage close on SMAP.
Isolated plateaus lowest, since each station overfits the single orbital patch
it sees. Centralized pools across stations, but
its second ground-to-central quantum hop adds noise to the training data and
slows convergence. \sol{} shares only classical parameters, so it gets the
pooled coverage without the hop, and its ESP-weighted aggregation down-weights
noisy backends, which FedAverage
lacks.

\begin{table}[t]
\centering
\caption{\textit{Detection quality of \sol{} under QPU calibration drift, as mean and standard deviation across runs.}}
\vspace{-4mm}
\label{tab:detection}
\scalebox{0.75}{
\begin{tabular}{cccccc}
\toprule
\rowcolor{headergray}
Dataset & Statistic & F1 & Precision & Recall & Accuracy \\
\midrule
 & Mean & 0.63 & 0.59 & 0.66 & 0.83 \\
\multirow{-2}{*}{OPS-SAT} & Standard Deviation (\%) & 4.3 & 6.8 & 2.4 & 2.9 \\
\midrule
 & Mean & 0.48 & 0.37 & 0.67 & 0.71 \\
\multirow{-2}{*}{SMAP} & Standard Deviation (\%) & 2.0 & 2.2 & 1.5 & 1.9 \\
\midrule 
 & Mean & 0.81 & 0.82 & 0.80 & 0.82 \\
\multirow{-2}{*}{MNIST} & Standard Deviation (\%) & 3.0 & 3.5 & 4.5 & 2.9 \\
\midrule
 & Mean & 0.83 & 0.83 & 0.84 & 0.84 \\
\multirow{-2}{*}{FashionMNIST} & Standard Deviation (\%) & 4.5 & 2.2 & 6.8 & 3.4 \\
\bottomrule
\end{tabular}
}
\vspace{-4mm}
\end{table}

\begin{figure}[t]
  \centering
  \includegraphics[width=\columnwidth]{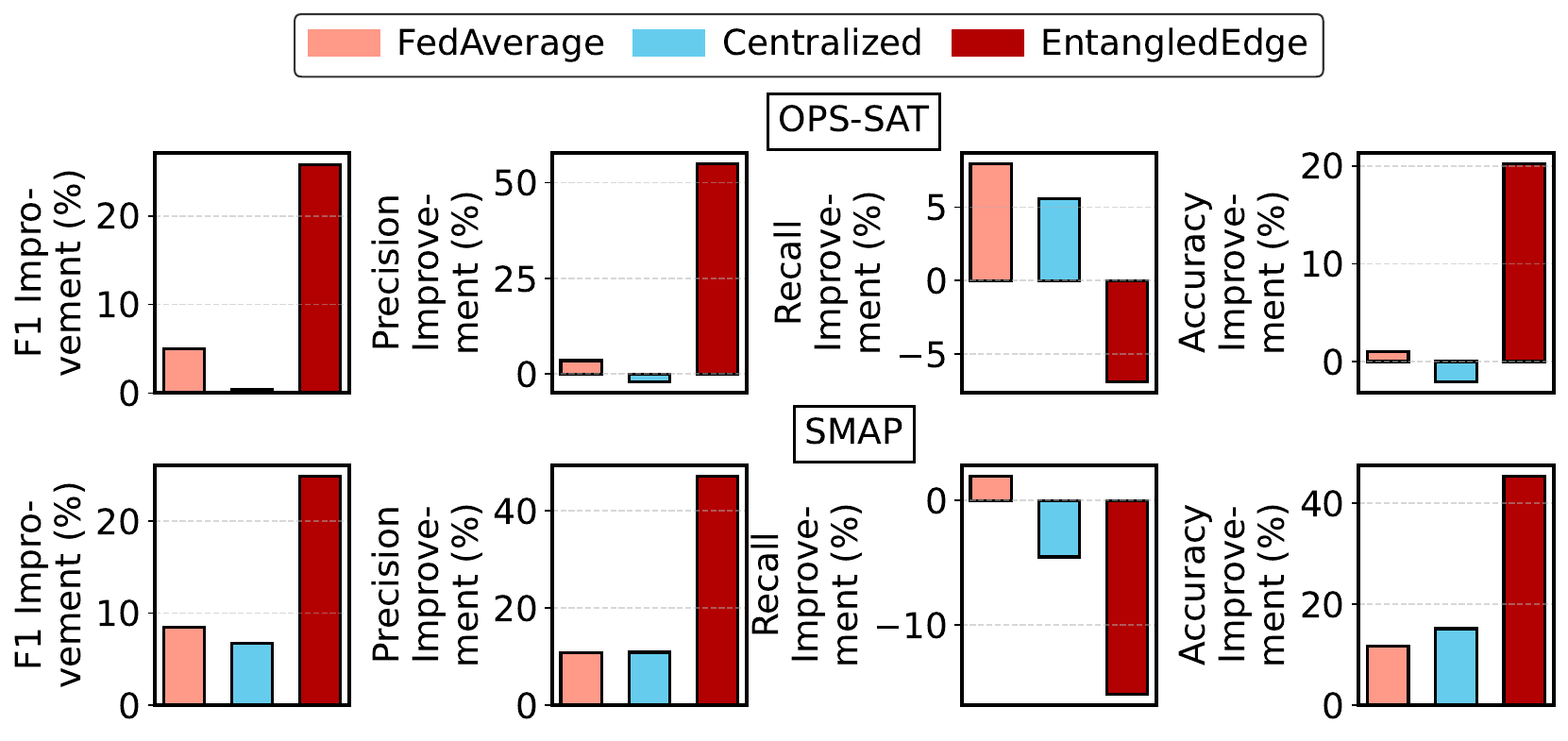}
  \vspace{-8mm}
\caption{\textit{\sol{} gains most in precision and dips in recall, but the precision gain outweighs it, so F1 (major metric) still leads (expressed as \% improvement over isolated).}}
  \label{fig:detection_improve}
  \vspace{-5mm}
\end{figure}

\sol{} also reaches the highest F1 in terms of raw value, 0.63 on OPS-SAT and
0.48 on SMAP (Table~\ref{tab:detection}) against 0.52 and 0.42 for FedAverage, and
leads on precision and accuracy. Each entry is a mean over 100 days of calibration drift
with a small standard deviation, so the ranking is stable. As improvement over isolated (Fig.~\ref{fig:detection_improve}), the
gain sits almost entirely in precision while recall drops a few points. The
reason is calibration. An overfit isolated model flags more candidates from a
miscalibrated distribution, so it trades precision for recall. \sol{} instead
routes each event to the bank member matched to its channel and judges the score
against a threshold that re-fits to the drifting distribution, which separates
normal from anomalous cleanly and suppresses false positives. On a sparse-anomaly
stream, this high-precision point is the useful one, and F1 settles in our
favor. The same table reports MNIST and FashionMNIST, where F1 stays above 0.80
under the identical setup, so the pipeline holds beyond the anomaly workload.

Every method runs online, so each event costs one inference and an SPSA update
per model it runs (Fig.~\ref{fig:flagship_system}, bottom, under the M/D/1/1
model at $T_{\text{infer}}=100$~ms). The single-model baselines run only the
large model at 0.30~s per event. \sol{} runs all three bank members
on every event, so each stays calibrated and ready when the channel drifts. Because the bank is capacity-diverse and the small and medium models
are cheaper, this readiness costs 0.57~s, $1.9\times$ the single-model figure and
well below the $3\times$ of three equal models. Centralized runs the same
single-model compute yet is slowest at 0.78~s, because every off-station event
pays a distance-scaled teleportation delay to the one central QPU, so latency is
set by the topology more than by the bank size.

\begin{figure}[t]
  \centering
  \includegraphics[width=\columnwidth]{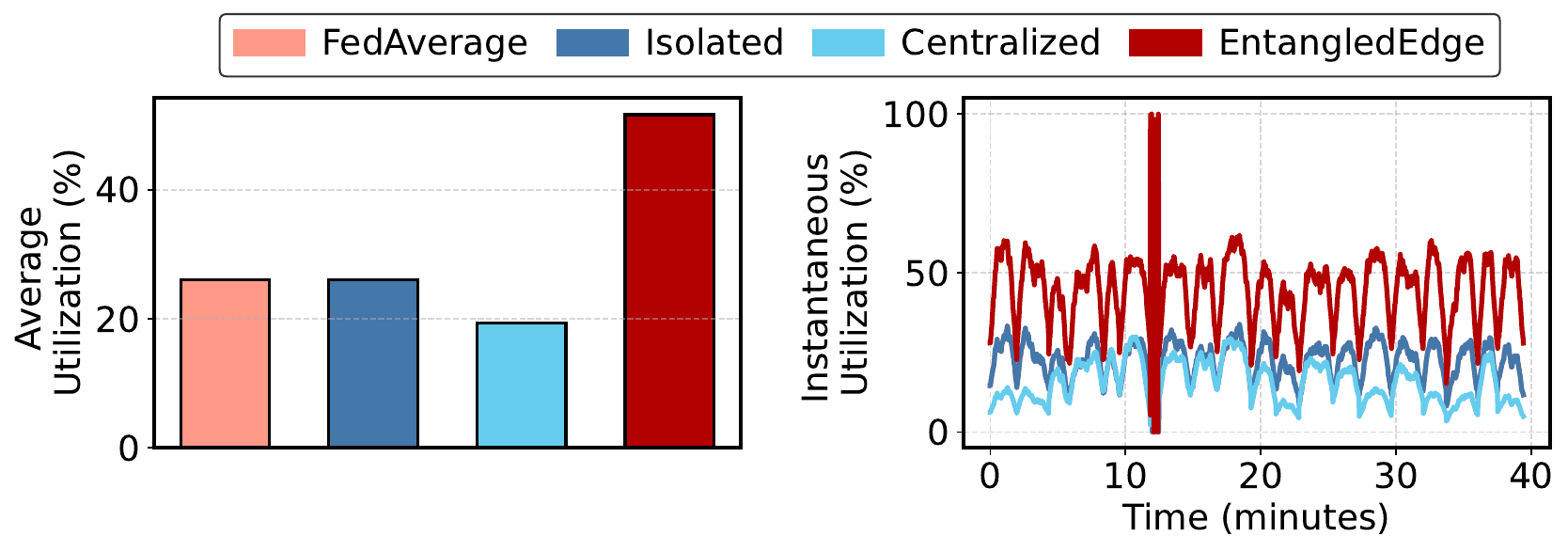}
  \vspace{-9mm}
\caption{\textit{\sol{} has the highest utilization due to training every bank member on each event, and centralized is lowest because one QPU does all the learning.}}
  \label{fig:utilization}
  \vspace{-4mm}
\end{figure}

\begin{figure}[t] 
  \centering
  \includegraphics[width=\columnwidth]{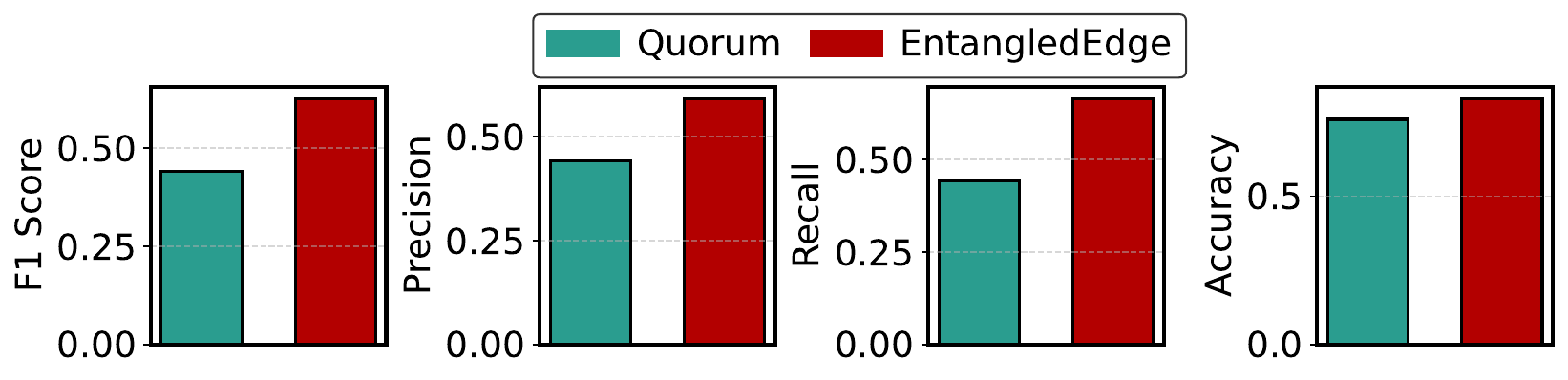}
  \vspace{-9mm}
\caption{\textit{\sol{} against Quorum on OPS-SAT data. \sol{} leads on every metric by adapting online to noise the zero-training competing technique cannot track.}}
  \label{fig:quorum}
  \vspace{-4mm}
\end{figure}
The same choice to train the whole bank locally raises active-window utilization
(Fig.~\ref{fig:utilization}), where \sol{} reaches 51.6\%, about twice that of FedAverage and isolated. \sol{} keeps every QPU working through its short
contact window, the efficient use of hardware that is otherwise idle most of the
orbit. Centralized instead routes all learning to one QPU, and its other three
only receive and forward states, so most of the fleet sits idle.

\sol{} also leads Quorum on all four metrics on OPS-SAT under matched channel and
QPU noise (Fig.~\ref{fig:quorum}, Sec.~\ref{sec:method:quorum}), with F1 at 0.63
against 0.44. Quorum reaches 0.62 F1 in an ideal noise-free run,
so its capacity is not what holds it back. It freezes random encoders and never
trains, so it cannot track the channel and QPU drift the deployment stream
carries. \sol{} trains online and refits its threshold to that drift, so its F1
holds while Quorum falls. A static detector cannot stay
calibrated in this orbital setting.

\vspace{-2mm}
\subsection{Robustness of \sol{}}
\label{sec:eval:robustness}

We stress \sol{} along three axes -- the QPU backend at each station, the QPU and
atmospheric noise levels, and the sensor noise itself. Every panel reports F1
improvement over the isolated per-station baseline on OPS-SAT.

\begin{figure}[t]
  \centering
  \includegraphics[width=\columnwidth]{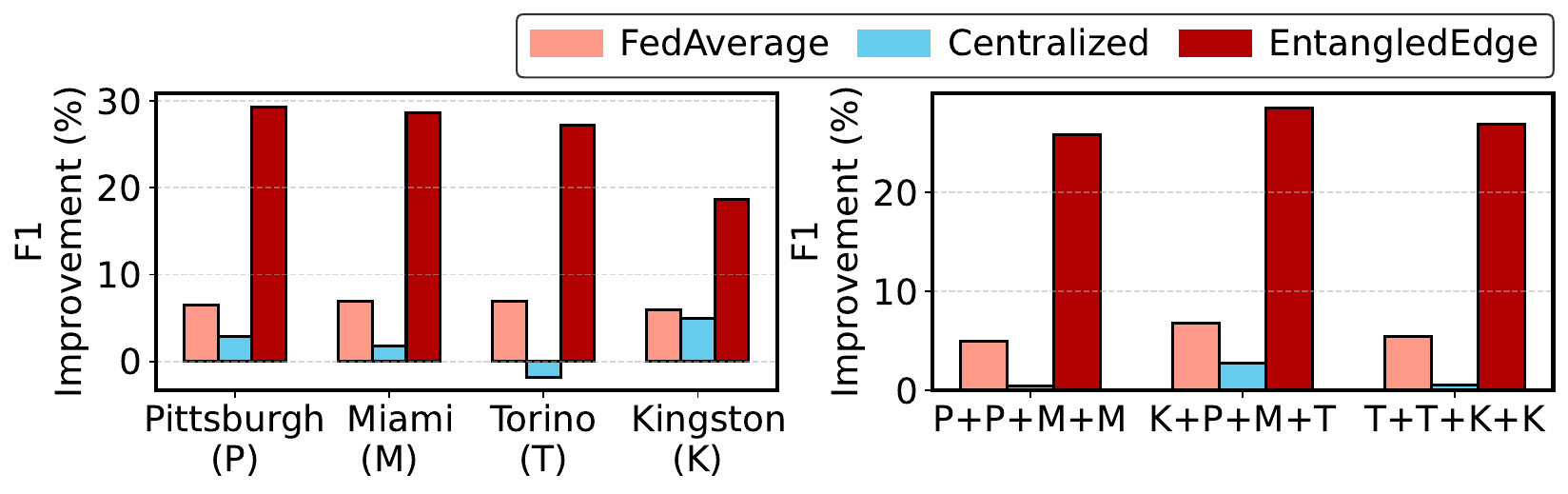}
  \vspace{-8mm}
  \caption{\textit{\sol{} improves F1 score over isolated baseline across backends on OPS-SAT, with every station on one backend (left) and a mix of backends across stations (right).}}
  \label{fig:backends}
  \vspace{-4mm}
\end{figure}

\sol{} results in the largest F1 improvement in every configuration
(Fig.~\ref{fig:backends}). The left figure runs all four stations on a single
shared backend, one backend at a time, and the right figure runs different
combinations of backends across the stations. FedAverage trails well behind in both,
and Centralized falls below the isolated baseline on the weakest backend. \sol{}
stays ahead because its ESP-weighted aggregation reads each backend's live
calibration and weights its update accordingly, so a weak or mismatched backend
contributes less and does not set the global model. The system is therefore robust across backends. 

\begin{figure}[t]
  \centering
  \includegraphics[width=\columnwidth]{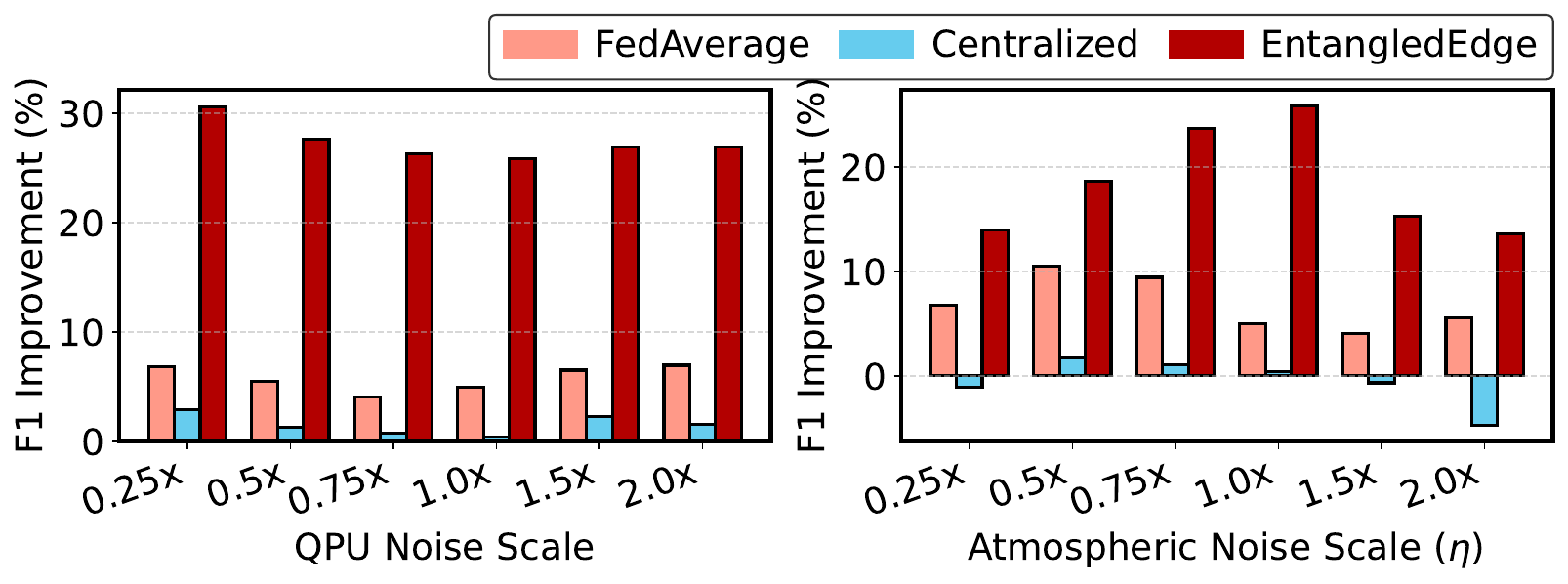}
  \vspace{-8mm}
  \caption{\textit{\sol{}'s F1 improvement over the isolated baseline under QPU noise scaling (left) and atmospheric noise scaling $\eta$ (right) on OPS-SAT.}}
  \label{fig:noise}
  \vspace{-4mm}
\end{figure}

We scale the QPU and atmospheric noise around the calibrated level
(Fig.~\ref{fig:noise}). Under QPU noise scaling, \sol{}'s F1 improvement is
roughly constant and largest at the low-noise end, because its
ESP weighting and federated averaging keep the noisier backends from degrading
the shared model. That low-noise end matters most, since it approximates the fault-tolerant (FTQC) hardware to come. \sol{} still
improves over isolated there, so its value does not depend on today's high noise.
Under atmospheric scaling, \sol{} shows the largest improvement at each severity
and remains positive across the range, because its router re-matches the model to
the channel as conditions change. Centralized is most affected on the channel,
since its second ground-to-central hop compounds atmospheric noise, falling below the isolated baseline at the harshest settings.

\begin{figure}[t]
  \centering
  \includegraphics[width=\columnwidth]{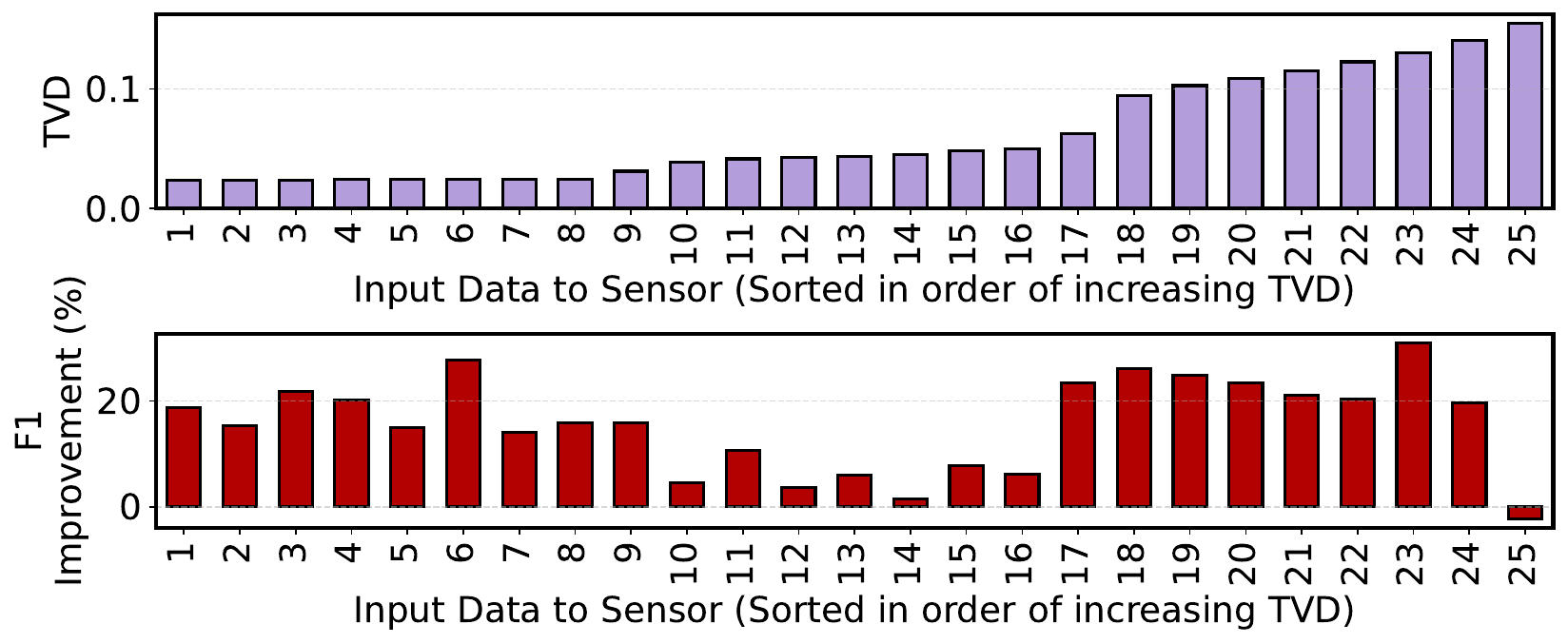}
  \vspace{-8mm}
\caption{\textit{\sol{} keeps improving F1 over the isolated baseline as sensor distortion grows (OPS-SAT). Chunks are ordered by mean TVD from the clean input (top), and the improvement stays positive across nearly all of them (bottom).}}
  \label{fig:sensor}
  \vspace{-4mm}
\end{figure}

The other experiments leave sensor noise out, since it perturbs every method's
input alike and does not change their ranking (Sec.~\ref{sec:method:sensor}). Here
we inject the real neutral-atom sensor noise into the state $\rho$, on top of the
channel and QPU noise the pipeline already applies. The top panel of
Fig.~\ref{fig:sensor} reports the total variation distance (TVD) between each
event's clean and sensor-perturbed input state, which isolates the sensor's own
contribution to the noise, and the sensor distorts different inputs by widely
varying amounts. We sort events by that distortion from light to heavy. \sol{}
still improves F1 over isolated by about 21\% overall and stays positive in nearly
every chunk across that range (bottom). The gain holds because \sol{}'s advantage
comes from matching capacity to the channel and sharing across stations, which the
input-side perturbation leaves intact.



\vspace{-2mm}
\subsection{Why does \sol{} Work ?}

\begin{figure}[t]
  \centering
  \includegraphics[width=\columnwidth]{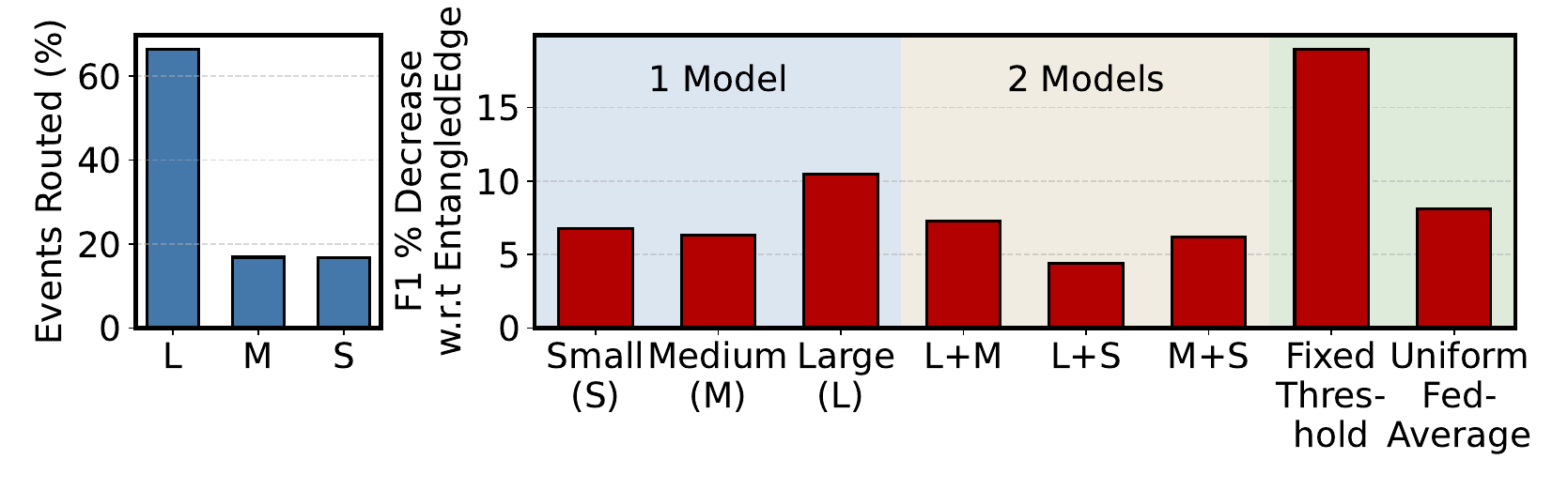}
  \vspace{-10mm}
\caption{\textit{Every design component matters (result for OPS-SAT data). The router uses all three capacities, and ablating the bank, the dynamic threshold, or ESP weighting lowers F1.}}
  \label{fig:ablation}
  \vspace{-4mm}
\end{figure}

We examine where \sol{}'s gains come from (Fig.~\ref{fig:ablation}). The router
spreads events across all three model sizes (left). The large model takes most
traffic, but the medium and small models carry a meaningful share, because
the router matches every event to its channel and the channel varies
(Sec.~\ref{sec:design:qaebank}).

To capture the benefit of each of \sol{}'s three main design components, we remove
one at a time and report the F1 drop against the full system
(Fig.~\ref{fig:ablation}, right). No single model or two-model pair matches the
full bank, and the large model alone is the worst single choice despite taking the
most traffic, because it suits the events routed to it but degrades on the noisier
ones the smaller models handle. Replacing the dynamic GMM-2 threshold
(Sec.~\ref{sec:design:qaebank}) with a fixed cutoff gives the largest drop,
because the score distribution drifts with the channel and only a re-fitting
threshold tracks it. Replacing ESP-weighted aggregation (Sec.~\ref{sec:design:esp})
with uniform FedAverage lets the noisier QPUs bias the global model. Each component
contributes to \sol{}'s effectiveness.

\subsection{Scaling of \sol{}}
\label{sec:eval:scaling}

\begin{figure}[t]
  \centering
  \includegraphics[width=\columnwidth]{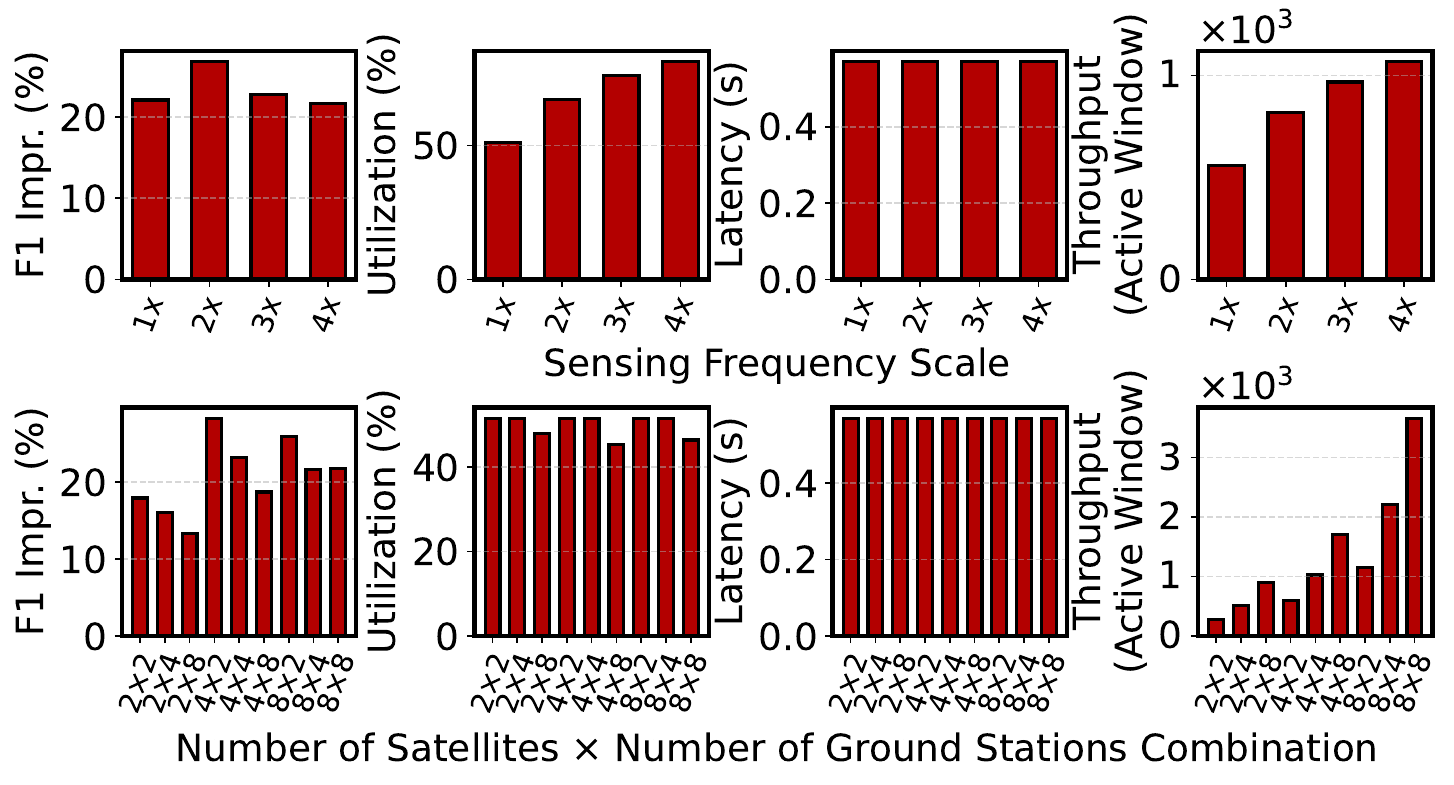}
  \vspace{-8mm}
\caption{\textit{\sol{} scales on both axes (OPS-SAT). A higher sensing rate raises utilization and throughput (top), and a larger fleet adds throughput in parallel (bottom), while per-event latency and F1 improvement stay roughly constant.}}
  \label{fig:scaling}
  \vspace{-4mm}
\end{figure}

We scale \sol{} along the sensing rate and the fleet size (Fig.~\ref{fig:scaling}).
Running every configuration at the full event scale is too
expensive, so the sensing-rate sweep uses a shortened simulation window, and the
fleet sweep grows its event volume with the configuration.

As the sensing rate increases from $1\times$ to $4\times$ (Fig.~\ref{fig:scaling},
top), more events arrive per visibility window, so QPU utilization increases from
51\% to 81\% and throughput roughly doubles. Per-event latency remains at 0.57~s,
since the M/D/1/1 service time does not depend on the arrival rate, and F1
improvement remains near 22 to 27\%. \sol{} handles the higher load by using
otherwise-idle QPU time, without increasing latency or reducing detection quality.

As the fleet size increases from $2\times2$ to $8\times8$ satellites and ground
stations (Fig.~\ref{fig:scaling}, bottom), throughput increases with it, from a few
hundred events to about 3,700 at $8\times8$, because each station serves its own
events in parallel. Per-event latency
remains at 0.57~s and utilization remains between 45 and 52\% across all nine
combinations, since the federated design has no central server to serialize the
work. F1 improvement remains positive throughout, from 13 to 28\%. Adding
satellites and stations increases throughput without raising per-event latency, so
\sol{} scales out.


Finally, we vary the large model's qubit count from 4 to 10 (medium and small fixed) on ibm\_pittsburgh hardware, and compare against the large model trained under the isolated baseline on OPS-SAT. Federation improves F1 at every size, and the gain grows with capacity over 6 to 10 qubits, from 8.4\% at 6 to 12.9\% at 8 and 18.3\% at 10, because a larger model better uses federation's cross-station data, with the largest effect at low channel noise. This trend generalizes beyond, showing \sol{} can scale to larger models.

\vspace{-2mm}
\subsection{Limitations}
\label{sec:eval:limits} 
\add{Orbital quantum sensors that emit their state into a quantum link are not yet deployed. We therefore apply Aquila-measured sensor noise to telemetry-encoded states and simulate the orbit-to-ground link (Table~\ref{tab:realvssim}). \sol{} is thus evaluated as a system architecture on a hybrid prototype, ahead of a deployable orbital stack.}

\begin{table}[t]
  \centering
  \caption{\textit{\sol{} against existing distributed systems.}}
  \vspace{-4mm}
  \label{tab:positioning}
  \scalebox{0.75}{
  \begin{tabular}{@{}>{\raggedright\arraybackslash}p{0.33\columnwidth}>{\raggedright\arraybackslash}p{0.29\columnwidth}>{\raggedright\arraybackslash}p{0.19\columnwidth}>{\raggedright\arraybackslash}p{0.40\columnwidth}@{}}
    \toprule
    \rowcolor{headergray}
    \textbf{Class} & \textbf{Crosses the link} & \textbf{Stored or re-sent?} & \textbf{\sol{} adds} \\
    \midrule
    Classical FL~\cite{mcmahan2017communication} & Parameters & Yes & Non-copyable quantum data \\
    \midrule
    Satellite edge, satellite FL~\cite{denby2020orbital, 10121575} & Sensed data; parameters & Yes, buffered on board & Unbufferable quantum data; heterogeneous QPUs \\
    \midrule
    Quantum networks~\cite{hermans2022qubit, lu2022micius}; DQC~\cite{monga2023quant, schon2024quant} & Prepared states; entangled photons & Re-prepared, re-sent & Sensed input that cannot be re-sent; a learner; model bank \\
    \midrule
    Distributed quantum sensing~\cite{malia2022distributed} & Measurement outcomes & Yes & State kept unmeasured across the link; heterogeneous QPUs \\
    \midrule
    Hybrid QML~\cite{cerezo2021variational, havlivcek2019supervised} & Nothing; one QPU & Re-prepared & Input over a noisy quantum link; model bank; many QPUs \\
    \midrule
    \sol{} & Sensed quantum state; parameters & No & --- \\
    \bottomrule
  \end{tabular}
  }
  \vspace{-4mm}
\end{table}

\vspace{-1mm}
\section{Related Works}
\label{sec:related}
\noindent\textbf{Orbital Edge Computing.} Orbital edge computing treats a satellite
constellation as a distributed system that computes onboard \cite{denby2020orbital}.
Later systems run ML inference in orbit \cite{denby2023kodan,zhang2024resource,liu2025spaceexit},
detection-driven capture \cite{cheng2024eagleeye}, and federated learning \cite{10121575, 10409275, lei2025joint, razmi2022ground}
and flown on an onboard accelerator \cite{giuffrida2021varphi}. Other work reduces
orbit-to-ground latency through distributed ground stations, task splitting, and
scheduling \cite{vasisht2021l2d2, tao2024known, lai2021orbitcast, cheng2025geoduck}.
Separate lines place compute in orbit as microdatacenters \cite{bleier2023space,
bleier2025architecting} and LEO cloud and serverless platforms \cite{bhosale2024krios,
pfandzelter2024komet, pfandzelter2025trabant}. All stay classical end-to-end, and none of these works
operate on quantum-sensed data that requires state preservation.

\vspace{1mm}

\noindent\textbf{Quantum Sensing, Networking, and Learning.} Quantum field sensing
surpasses the classical limit in magnetometry, gravimetry, and time-frequency
metrology \cite{RevModPhys.89.035002, pezze2018quantum}, with advances on atomic
and optical platforms \cite{cassens2025entanglement, yang2025clock, wu2023quantum, RevModPhys.89.035002,jebraeilli2026stqs, crawford2021quantum, aslam2023quantum, bass2024quantum}.
Quantum networks distribute entanglement and transfer states \cite{wehner2018quantum},
with teleportation \cite{valivarthi2016quantum, shen2023hertz} and entanglement
distribution \cite{stolk2024metropolitan, wengerowsky2019entanglement} over fiber
and multi-node testbeds \cite{pompili2021realization}. Satellites reach ground stations
\cite{lu2022micius, yin2017satellite} under studied schedules \cite{maule2024fair,
williams2024scalable, panigrahy2022optimal, hossain2025opportunistic, dikshit2025scheduling}\add{, and been proposed to entangle ground quantum sensors \cite{khalid2024integrated}},
and fiber testbeds target distributed quantum computing \cite{monga2023quant,
schon2024quant}. For learning, variational classifiers run on quantum hardware
\cite{havlivcek2019supervised, huang2022quantum}, through a hybrid
quantum-classical loop \cite{cerezo2021variational} whose parameters extend to
quantum federated learning \cite{chen2021federated, han2025layerwise, park2025quantum,
pokharel2025differentially, addo2025federated}\add{, including over satellite--ground links \cite{park2024dynamic}. Quantum learning has also been applied to LEO communication signals \cite{tong2026hybrid} and Earth-observation imagery \cite{miroszewski2026quo}. }\add{None of these works learns from an orbital sensed state that stays unmeasured. What they move across a link is classical data, model parameters, prepared states, or entanglement.}

\vspace{1mm}

\noindent\add{\textbf{Why existing remedies do not transfer?} Satellite edge computing, teleportation, and federated learning recover from a missed contact or a downed station by storing data, re-sending it, or holding it until it can be labeled~\cite{denby2020orbital, mcmahan2017communication, chehimi2022quantum} (Table~\ref{tab:positioning}). A sensed quantum state allows none of these. A missed pass loses the state, a failed transfer cannot be repeated, and labeling destroys it. Because these mechanisms do not carry over, link noise reaches the model directly, which is why we need the model bank and ESP-weighted aggregation.}


\vspace{-2mm}
\section{Conclusion}

\sol{} is the first \add{system architecture} to unify orbital quantum
sensing and learning. Its data plane stays quantum from the
orbital sensor to ground inference, and it improves detection quality, maintains high system utilization, and scales out
across fleets without raising per-event latency.

\vspace{2mm}

\noindent\add{\textbf{Acknowledgments.} We thank our shepherd and the reviewers for their feedback. This research used resources of the National Energy Research Scientific Computing Center (NERSC), a DOE Office of Science User Facility supported under Contract No.~DE-AC02-05CH11231, using NERSC award DDR-ERCAP0037923. This work was supported by the Kahlert School of Computing and the SCI Institute at the University of Utah. Rice University and DOE Quantum Testbed Finder Award DE-SC0024301 supported the Rice authors.}

% ...

\bibliographystyle{ACM-Reference-Format}
\bibliography{references}


\begin{thebibliography}{96}


\ifx \showCODEN    \undefined \def \showCODEN     #1{\unskip}     \fi
\ifx \showISBNx    \undefined \def \showISBNx     #1{\unskip}     \fi
\ifx \showISBNxiii \undefined \def \showISBNxiii  #1{\unskip}     \fi
\ifx \showISSN     \undefined \def \showISSN      #1{\unskip}     \fi
\ifx \showLCCN     \undefined \def \showLCCN      #1{\unskip}     \fi
\ifx \shownote     \undefined \def \shownote      #1{#1}          \fi
\ifx \showarticletitle \undefined \def \showarticletitle #1{#1}   \fi
\ifx \showURL      \undefined \def \showURL       {\relax}        \fi
\providecommand\bibfield[2]{#2}
\providecommand\bibinfo[2]{#2}
\providecommand\natexlab[1]{#1}
\providecommand\showeprint[2][]{arXiv:#2}

\bibitem[ibm({[n.\,d.]})]%
        {ibmEfficientSU2latest}
 \bibinfo{year}{[n.\,d.]}\natexlab{}.
\newblock \bibinfo{title}{{E}fficient{S}{U}2 (latest version) | {I}{B}{M} {Q}uantum {D}ocumentation --- quantum.cloud.ibm.com}.
\newblock \bibinfo{howpublished}{\url{https://quantum.cloud.ibm.com/docs/en/api/qiskit/qiskit.circuit.library.{E}fficient{S}{U}2}}.
\newblock


\bibitem[Addo et~al\mbox{.}(2025)]%
        {addo2025federated}
\bibfield{author}{\bibinfo{person}{Kwabena Addo}, \bibinfo{person}{Musasa Kabeya}, {and} \bibinfo{person}{Evans~Eshiemogie Ojo}.} \bibinfo{year}{2025}\natexlab{}.
\newblock \showarticletitle{Federated quantum machine learning for distributed cybersecurity in multi-agent energy systems}.
\newblock \bibinfo{journal}{\emph{Energies}} \bibinfo{volume}{18}, \bibinfo{number}{20} (\bibinfo{year}{2025}), \bibinfo{pages}{5418}.
\newblock


\bibitem[Andrews and Phillips(2005)]%
        {andrews2005laser}
\bibfield{author}{\bibinfo{person}{Larry~C Andrews} {and} \bibinfo{person}{Ronald~L Phillips}.} \bibinfo{year}{2005}\natexlab{}.
\newblock \showarticletitle{Laser beam propagation through random media}.
\newblock \bibinfo{journal}{\emph{Laser Beam Propagation Through Random Media: Second Edition}} (\bibinfo{year}{2005}).
\newblock


\bibitem[Aslam et~al\mbox{.}(2023)]%
        {aslam2023quantum}
\bibfield{author}{\bibinfo{person}{Nabeel Aslam}, \bibinfo{person}{Hengyun Zhou}, \bibinfo{person}{Elana~K Urbach}, \bibinfo{person}{Matthew~J Turner}, \bibinfo{person}{Ronald~L Walsworth}, \bibinfo{person}{Mikhail~D Lukin}, {and} \bibinfo{person}{Hongkun Park}.} \bibinfo{year}{2023}\natexlab{}.
\newblock \showarticletitle{Quantum sensors for biomedical applications}.
\newblock \bibinfo{journal}{\emph{Nature Reviews Physics}} \bibinfo{volume}{5}, \bibinfo{number}{3} (\bibinfo{year}{2023}), \bibinfo{pages}{157--169}.
\newblock


\bibitem[Bass and Doser(2024)]%
        {bass2024quantum}
\bibfield{author}{\bibinfo{person}{Steven~D Bass} {and} \bibinfo{person}{Michael Doser}.} \bibinfo{year}{2024}\natexlab{}.
\newblock \showarticletitle{Quantum sensing for particle physics}.
\newblock \bibinfo{journal}{\emph{Nature Reviews Physics}} \bibinfo{volume}{6}, \bibinfo{number}{5} (\bibinfo{year}{2024}), \bibinfo{pages}{329--339}.
\newblock


\bibitem[Behera and Sinha(2024)]%
        {behera2024estimating}
\bibfield{author}{\bibinfo{person}{Satya~Ranjan Behera} {and} \bibinfo{person}{Urbasi Sinha}.} \bibinfo{year}{2024}\natexlab{}.
\newblock \showarticletitle{Estimating the link budget of satellite-based Quantum Key Distribution (QKD) for uplink transmission through the atmosphere}.
\newblock \bibinfo{journal}{\emph{EPJ Quantum Technology}} \bibinfo{volume}{11}, \bibinfo{number}{1} (\bibinfo{year}{2024}), \bibinfo{pages}{66}.
\newblock


\bibitem[Bhosale et~al\mbox{.}(2024)]%
        {bhosale2024krios}
\bibfield{author}{\bibinfo{person}{Vaibhav Bhosale}, \bibinfo{person}{Ada Gavrilovska}, {and} \bibinfo{person}{Ketan Bhardwaj}.} \bibinfo{year}{2024}\natexlab{}.
\newblock \showarticletitle{Krios: Scheduling abstractions and mechanisms for enabling a LEO compute cloud}. In \bibinfo{booktitle}{\emph{Proceedings of the 2024 ACM Symposium on Cloud Computing}}. \bibinfo{pages}{322--340}.
\newblock


\bibitem[Bleier et~al\mbox{.}(2025)]%
        {bleier2025architecting}
\bibfield{author}{\bibinfo{person}{Nathan Bleier}, \bibinfo{person}{Rick Eason}, \bibinfo{person}{Michael Lembeck}, {and} \bibinfo{person}{Rakesh Kumar}.} \bibinfo{year}{2025}\natexlab{}.
\newblock \showarticletitle{Architecting space microdatacenters: A system-level approach}. In \bibinfo{booktitle}{\emph{2025 IEEE International Symposium on High Performance Computer Architecture (HPCA)}}. IEEE, \bibinfo{pages}{1304--1319}.
\newblock


\bibitem[Bleier et~al\mbox{.}(2023)]%
        {bleier2023space}
\bibfield{author}{\bibinfo{person}{Nathaniel Bleier}, \bibinfo{person}{Muhammad~Husnain Mubarik}, \bibinfo{person}{Gary~R Swenson}, {and} \bibinfo{person}{Rakesh Kumar}.} \bibinfo{year}{2023}\natexlab{}.
\newblock \showarticletitle{Space microdatacenters}. In \bibinfo{booktitle}{\emph{Proceedings of the 56th Annual IEEE/ACM International Symposium on Microarchitecture}}. \bibinfo{pages}{900--915}.
\newblock


\bibitem[Bourgoin et~al\mbox{.}(2013)]%
        {bourgoin2013comprehensive}
\bibfield{author}{\bibinfo{person}{Jean-Philippe Bourgoin}, \bibinfo{person}{Evan Meyer-Scott}, \bibinfo{person}{Brendon~L Higgins}, \bibinfo{person}{B Helou}, \bibinfo{person}{Chris Erven}, \bibinfo{person}{Hannes Huebel}, \bibinfo{person}{B Kumar}, \bibinfo{person}{D Hudson}, \bibinfo{person}{Ian D'Souza}, \bibinfo{person}{Ralph Girard}, {et~al\mbox{.}}} \bibinfo{year}{2013}\natexlab{}.
\newblock \showarticletitle{A comprehensive design and performance analysis of low Earth orbit satellite quantum communication}.
\newblock \bibinfo{journal}{\emph{New Journal of Physics}} \bibinfo{volume}{15}, \bibinfo{number}{2} (\bibinfo{year}{2013}), \bibinfo{pages}{023006}.
\newblock


\bibitem[Bu{\v{z}}ek and Hillery(1996)]%
        {buvzek1996quantum}
\bibfield{author}{\bibinfo{person}{Vladimir Bu{\v{z}}ek} {and} \bibinfo{person}{Mark Hillery}.} \bibinfo{year}{1996}\natexlab{}.
\newblock \showarticletitle{Quantum copying: Beyond the no-cloning theorem}.
\newblock \bibinfo{journal}{\emph{Physical Review A}} \bibinfo{volume}{54}, \bibinfo{number}{3} (\bibinfo{year}{1996}), \bibinfo{pages}{1844}.
\newblock


\bibitem[Cassens et~al\mbox{.}(2025)]%
        {cassens2025entanglement}
\bibfield{author}{\bibinfo{person}{Christophe Cassens}, \bibinfo{person}{Bernd Meyer-Hoppe}, \bibinfo{person}{Ernst Rasel}, {and} \bibinfo{person}{Carsten Klempt}.} \bibinfo{year}{2025}\natexlab{}.
\newblock \showarticletitle{Entanglement-enhanced atomic gravimeter}.
\newblock \bibinfo{journal}{\emph{Physical Review X}} \bibinfo{volume}{15}, \bibinfo{number}{1} (\bibinfo{year}{2025}), \bibinfo{pages}{011029}.
\newblock


\bibitem[Cerezo et~al\mbox{.}(2021)]%
        {cerezo2021variational}
\bibfield{author}{\bibinfo{person}{Marco Cerezo}, \bibinfo{person}{Andrew Arrasmith}, \bibinfo{person}{Ryan Babbush}, \bibinfo{person}{Simon~C Benjamin}, \bibinfo{person}{Suguru Endo}, \bibinfo{person}{Keisuke Fujii}, \bibinfo{person}{Jarrod~R McClean}, \bibinfo{person}{Kosuke Mitarai}, \bibinfo{person}{Xiao Yuan}, \bibinfo{person}{Lukasz Cincio}, {et~al\mbox{.}}} \bibinfo{year}{2021}\natexlab{}.
\newblock \showarticletitle{Variational quantum algorithms}.
\newblock \bibinfo{journal}{\emph{Nature Reviews Physics}} \bibinfo{volume}{3}, \bibinfo{number}{9} (\bibinfo{year}{2021}), \bibinfo{pages}{625--644}.
\newblock


\bibitem[Chehimi and Saad(2022)]%
        {chehimi2022quantum}
\bibfield{author}{\bibinfo{person}{Mahdi Chehimi} {and} \bibinfo{person}{Walid Saad}.} \bibinfo{year}{2022}\natexlab{}.
\newblock \showarticletitle{Quantum federated learning with quantum data}. In \bibinfo{booktitle}{\emph{ICASSP 2022-2022 IEEE International Conference on Acoustics, Speech and Signal Processing (ICASSP)}}. IEEE, \bibinfo{pages}{8617--8621}.
\newblock


\bibitem[Chen and Yoo(2021)]%
        {chen2021federated}
\bibfield{author}{\bibinfo{person}{Samuel Yen-Chi Chen} {and} \bibinfo{person}{Shinjae Yoo}.} \bibinfo{year}{2021}\natexlab{}.
\newblock \showarticletitle{Federated quantum machine learning}.
\newblock \bibinfo{journal}{\emph{Entropy}} \bibinfo{volume}{23}, \bibinfo{number}{4} (\bibinfo{year}{2021}), \bibinfo{pages}{460}.
\newblock


\bibitem[Cheng et~al\mbox{.}(2024)]%
        {cheng2024eagleeye}
\bibfield{author}{\bibinfo{person}{Zhuo Cheng}, \bibinfo{person}{Bradley Denby}, \bibinfo{person}{Kyle McCleary}, {and} \bibinfo{person}{Brandon Lucia}.} \bibinfo{year}{2024}\natexlab{}.
\newblock \showarticletitle{EagleEye: Nanosatellite constellation design for high-coverage, high-resolution sensing}. In \bibinfo{booktitle}{\emph{Proceedings of the 29th ACM International Conference on Architectural Support for Programming Languages and Operating Systems, Volume 1}}. \bibinfo{pages}{117--132}.
\newblock


\bibitem[Cheng and Lucia(2025)]%
        {cheng2025geoduck}
\bibfield{author}{\bibinfo{person}{Zhuo Cheng} {and} \bibinfo{person}{Brandon Lucia}.} \bibinfo{year}{2025}\natexlab{}.
\newblock \showarticletitle{Geoduck: Nanosatellite Constellation Scheduling for Low Latency Event Detection}. In \bibinfo{booktitle}{\emph{Proceedings of the 23rd ACM Conference on Embedded Networked Sensor Systems}}. \bibinfo{pages}{547--559}.
\newblock


\bibitem[Crawford et~al\mbox{.}(2021)]%
        {crawford2021quantum}
\bibfield{author}{\bibinfo{person}{Scott~E Crawford}, \bibinfo{person}{Roman~A Shugayev}, \bibinfo{person}{Hari~P Paudel}, \bibinfo{person}{Ping Lu}, \bibinfo{person}{Madhava Syamlal}, \bibinfo{person}{Paul~R Ohodnicki}, \bibinfo{person}{Benjamin Chorpening}, \bibinfo{person}{Randall Gentry}, {and} \bibinfo{person}{Yuhua Duan}.} \bibinfo{year}{2021}\natexlab{}.
\newblock \showarticletitle{Quantum sensing for energy applications: Review and perspective}.
\newblock \bibinfo{journal}{\emph{Advanced Quantum Technologies}} \bibinfo{volume}{4}, \bibinfo{number}{8} (\bibinfo{year}{2021}), \bibinfo{pages}{2100049}.
\newblock


\bibitem[Davis and Goadrich(2006)]%
        {davis2006relationship}
\bibfield{author}{\bibinfo{person}{Jesse Davis} {and} \bibinfo{person}{Mark Goadrich}.} \bibinfo{year}{2006}\natexlab{}.
\newblock \showarticletitle{The relationship between Precision-Recall and ROC curves}. In \bibinfo{booktitle}{\emph{Proceedings of the 23rd international conference on Machine learning}}. \bibinfo{pages}{233--240}.
\newblock


\bibitem[Degen et~al\mbox{.}(2017)]%
        {RevModPhys.89.035002}
\bibfield{author}{\bibinfo{person}{C.~L. Degen}, \bibinfo{person}{F. Reinhard}, {and} \bibinfo{person}{P. Cappellaro}.} \bibinfo{year}{2017}\natexlab{}.
\newblock \showarticletitle{Quantum sensing}.
\newblock \bibinfo{journal}{\emph{Rev. Mod. Phys.}}  \bibinfo{volume}{89} (\bibinfo{date}{Jul} \bibinfo{year}{2017}), \bibinfo{pages}{035002}.
\newblock
Issue 3.
\href{https://doi.org/10.1103/RevModPhys.89.035002}{doi:\nolinkurl{10.1103/RevModPhys.89.035002}}


\bibitem[Denby et~al\mbox{.}(2023)]%
        {denby2023kodan}
\bibfield{author}{\bibinfo{person}{Bradley Denby}, \bibinfo{person}{Krishna Chintalapudi}, \bibinfo{person}{Ranveer Chandra}, \bibinfo{person}{Brandon Lucia}, {and} \bibinfo{person}{Shadi Noghabi}.} \bibinfo{year}{2023}\natexlab{}.
\newblock \showarticletitle{Kodan: Addressing the computational bottleneck in space}. In \bibinfo{booktitle}{\emph{Proceedings of the 28th ACM International Conference on Architectural Support for Programming Languages and Operating Systems, Volume 3}}. \bibinfo{pages}{392--403}.
\newblock


\bibitem[Denby and Lucia(2020)]%
        {denby2020orbital}
\bibfield{author}{\bibinfo{person}{Bradley Denby} {and} \bibinfo{person}{Brandon Lucia}.} \bibinfo{year}{2020}\natexlab{}.
\newblock \showarticletitle{Orbital edge computing: Nanosatellite constellations as a new class of computer system}. In \bibinfo{booktitle}{\emph{Proceedings of the Twenty-Fifth International Conference on Architectural Support for Programming Languages and Operating Systems}}. \bibinfo{pages}{939--954}.
\newblock


\bibitem[Dikshit et~al\mbox{.}(2025)]%
        {dikshit2025scheduling}
\bibfield{author}{\bibinfo{person}{Ashutosh~Jayant Dikshit}, \bibinfo{person}{Naga~Lakshmi Anipeddi}, \bibinfo{person}{Prajit Dhara}, \bibinfo{person}{Saikat Guha}, \bibinfo{person}{Deirdre Kilbane}, \bibinfo{person}{Leandros Tassiulas}, \bibinfo{person}{Don Towsley}, {and} \bibinfo{person}{Nitish~K Panigrahy}.} \bibinfo{year}{2025}\natexlab{}.
\newblock \showarticletitle{Scheduling in Quantum Satellite Networks: Fairness and Performance Optimization}.
\newblock \bibinfo{journal}{\emph{arXiv preprint arXiv:2512.07108}} (\bibinfo{year}{2025}).
\newblock


\bibitem[Filip(2024)]%
        {filip2024fighting}
\bibfield{author}{\bibinfo{person}{Maria-Andreea Filip}.} \bibinfo{year}{2024}\natexlab{}.
\newblock \showarticletitle{Fighting noise with noise: a stochastic projective quantum eigensolver}.
\newblock \bibinfo{journal}{\emph{Journal of Chemical Theory and Computation}} \bibinfo{volume}{20}, \bibinfo{number}{14} (\bibinfo{year}{2024}), \bibinfo{pages}{5964--5981}.
\newblock


\bibitem[Ghalaii and Pirandola(2022)]%
        {ghalaii2022quantum}
\bibfield{author}{\bibinfo{person}{Masoud Ghalaii} {and} \bibinfo{person}{Stefano Pirandola}.} \bibinfo{year}{2022}\natexlab{}.
\newblock \showarticletitle{Quantum communications in a moderate-to-strong turbulent space}.
\newblock \bibinfo{journal}{\emph{Communications Physics}} \bibinfo{volume}{5}, \bibinfo{number}{1} (\bibinfo{year}{2022}), \bibinfo{pages}{38}.
\newblock


\bibitem[Giovannetti et~al\mbox{.}(2006)]%
        {giovannetti2006quantum}
\bibfield{author}{\bibinfo{person}{Vittorio Giovannetti}, \bibinfo{person}{Seth Lloyd}, {and} \bibinfo{person}{Lorenzo Maccone}.} \bibinfo{year}{2006}\natexlab{}.
\newblock \showarticletitle{Quantum metrology}.
\newblock \bibinfo{journal}{\emph{Physical review letters}} \bibinfo{volume}{96}, \bibinfo{number}{1} (\bibinfo{year}{2006}), \bibinfo{pages}{010401}.
\newblock


\bibitem[Giuffrida et~al\mbox{.}(2021)]%
        {giuffrida2021varphi}
\bibfield{author}{\bibinfo{person}{Gianluca Giuffrida}, \bibinfo{person}{Luca Fanucci}, \bibinfo{person}{Gabriele Meoni}, \bibinfo{person}{Matej Bati{\v{c}}}, \bibinfo{person}{L{\'e}onie Buckley}, \bibinfo{person}{Aubrey Dunne}, \bibinfo{person}{Chris Van~Dijk}, \bibinfo{person}{Marco Esposito}, \bibinfo{person}{John Hefele}, \bibinfo{person}{Nathan Vercruyssen}, {et~al\mbox{.}}} \bibinfo{year}{2021}\natexlab{}.
\newblock \showarticletitle{The $\Phi$-Sat-1 mission: The first on-board deep neural network demonstrator for satellite earth observation}.
\newblock \bibinfo{journal}{\emph{IEEE Transactions on Geoscience and Remote Sensing}}  \bibinfo{volume}{60} (\bibinfo{year}{2021}), \bibinfo{pages}{1--14}.
\newblock


\bibitem[Gonzalez-Raya et~al\mbox{.}(2024)]%
        {gonzalez2024satellite}
\bibfield{author}{\bibinfo{person}{Tasio Gonzalez-Raya}, \bibinfo{person}{Stefano Pirandola}, {and} \bibinfo{person}{Mikel Sanz}.} \bibinfo{year}{2024}\natexlab{}.
\newblock \showarticletitle{Satellite-based entanglement distribution and quantum teleportation with continuous variables}.
\newblock \bibinfo{journal}{\emph{Communications Physics}} \bibinfo{volume}{7}, \bibinfo{number}{1} (\bibinfo{year}{2024}), \bibinfo{pages}{126}.
\newblock


\bibitem[Han et~al\mbox{.}(2025)]%
        {han2025layerwise}
\bibfield{author}{\bibinfo{person}{Jason Han}, \bibinfo{person}{Nicholas~S DiBrita}, \bibinfo{person}{Daniel Leeds}, \bibinfo{person}{Jianqiang Li}, \bibinfo{person}{Jason Ludmir}, {and} \bibinfo{person}{Tirthak Patel}.} \bibinfo{year}{2025}\natexlab{}.
\newblock \showarticletitle{Layerwise federated learning for heterogeneous quantum clients using quorus}.
\newblock \bibinfo{journal}{\emph{arXiv preprint arXiv:2510.06228}} (\bibinfo{year}{2025}).
\newblock


\bibitem[Han et~al\mbox{.}(2020)]%
        {han2020polarization}
\bibfield{author}{\bibinfo{person}{Xuan Han}, \bibinfo{person}{Hai-Lin Yong}, \bibinfo{person}{Ping Xu}, \bibinfo{person}{Kui-Xing Yang}, \bibinfo{person}{Shuang-Lin Li}, \bibinfo{person}{Wei-Yang Wang}, \bibinfo{person}{Hua-Jian Xue}, \bibinfo{person}{Feng-Zhi Li}, \bibinfo{person}{Ji-Gang Ren}, \bibinfo{person}{Cheng-Zhi Peng}, {et~al\mbox{.}}} \bibinfo{year}{2020}\natexlab{}.
\newblock \showarticletitle{Polarization design for ground-to-satellite quantum entanglement distribution}.
\newblock \bibinfo{journal}{\emph{Optics Express}} \bibinfo{volume}{28}, \bibinfo{number}{1} (\bibinfo{year}{2020}), \bibinfo{pages}{369--378}.
\newblock


\bibitem[Havl{\'\i}{\v{c}}ek et~al\mbox{.}(2019)]%
        {havlivcek2019supervised}
\bibfield{author}{\bibinfo{person}{Vojt{\v{e}}ch Havl{\'\i}{\v{c}}ek}, \bibinfo{person}{Antonio~D C{\'o}rcoles}, \bibinfo{person}{Kristan Temme}, \bibinfo{person}{Aram~W Harrow}, \bibinfo{person}{Abhinav Kandala}, \bibinfo{person}{Jerry~M Chow}, {and} \bibinfo{person}{Jay~M Gambetta}.} \bibinfo{year}{2019}\natexlab{}.
\newblock \showarticletitle{Supervised learning with quantum-enhanced feature spaces}.
\newblock \bibinfo{journal}{\emph{Nature}} \bibinfo{volume}{567}, \bibinfo{number}{7747} (\bibinfo{year}{2019}), \bibinfo{pages}{209--212}.
\newblock


\bibitem[Hermans et~al\mbox{.}(2022)]%
        {hermans2022qubit}
\bibfield{author}{\bibinfo{person}{SLN Hermans}, \bibinfo{person}{Matteo Pompili}, \bibinfo{person}{HKC Beukers}, \bibinfo{person}{Simon Baier}, \bibinfo{person}{Johannes Borregaard}, {and} \bibinfo{person}{Ronald Hanson}.} \bibinfo{year}{2022}\natexlab{}.
\newblock \showarticletitle{Qubit teleportation between non-neighbouring nodes in a quantum network}.
\newblock \bibinfo{journal}{\emph{Nature}} \bibinfo{volume}{605}, \bibinfo{number}{7911} (\bibinfo{year}{2022}), \bibinfo{pages}{663--668}.
\newblock


\bibitem[Hossain et~al\mbox{.}(2025)]%
        {hossain2025opportunistic}
\bibfield{author}{\bibinfo{person}{Md~Zakir Hossain}, \bibinfo{person}{Nitish~K Panigrahy}, \bibinfo{person}{Walter~O Krawec}, \bibinfo{person}{Don Towsley}, {and} \bibinfo{person}{Bing Wang}.} \bibinfo{year}{2025}\natexlab{}.
\newblock \showarticletitle{Opportunistic Scheduling for Single-downlink Satellite-based Quantum Key Distribution}.
\newblock \bibinfo{journal}{\emph{arXiv preprint arXiv:2512.12514}} (\bibinfo{year}{2025}).
\newblock


\bibitem[Huang et~al\mbox{.}(2022)]%
        {huang2022quantum}
\bibfield{author}{\bibinfo{person}{Hsin-Yuan Huang}, \bibinfo{person}{Michael Broughton}, \bibinfo{person}{Jordan Cotler}, \bibinfo{person}{Sitan Chen}, \bibinfo{person}{Jerry Li}, \bibinfo{person}{Masoud Mohseni}, \bibinfo{person}{Hartmut Neven}, \bibinfo{person}{Ryan Babbush}, \bibinfo{person}{Richard Kueng}, \bibinfo{person}{John Preskill}, {et~al\mbox{.}}} \bibinfo{year}{2022}\natexlab{}.
\newblock \showarticletitle{Quantum advantage in learning from experiments}.
\newblock \bibinfo{journal}{\emph{Science}} \bibinfo{volume}{376}, \bibinfo{number}{6598} (\bibinfo{year}{2022}), \bibinfo{pages}{1182--1186}.
\newblock


\bibitem[Huang et~al\mbox{.}(2021)]%
        {huang2021power}
\bibfield{author}{\bibinfo{person}{Hsin-Yuan Huang}, \bibinfo{person}{Michael Broughton}, \bibinfo{person}{Masoud Mohseni}, \bibinfo{person}{Ryan Babbush}, \bibinfo{person}{Sergio Boixo}, \bibinfo{person}{Hartmut Neven}, {and} \bibinfo{person}{Jarrod~R McClean}.} \bibinfo{year}{2021}\natexlab{}.
\newblock \showarticletitle{Power of data in quantum machine learning}.
\newblock \bibinfo{journal}{\emph{Nature communications}} \bibinfo{volume}{12}, \bibinfo{number}{1} (\bibinfo{year}{2021}), \bibinfo{pages}{2631}.
\newblock


\bibitem[Hundman et~al\mbox{.}(2018)]%
        {10.1145/3219819.3219845}
\bibfield{author}{\bibinfo{person}{Kyle Hundman}, \bibinfo{person}{Valentino Constantinou}, \bibinfo{person}{Christopher Laporte}, \bibinfo{person}{Ian Colwell}, {and} \bibinfo{person}{Tom Soderstrom}.} \bibinfo{year}{2018}\natexlab{}.
\newblock \showarticletitle{Detecting Spacecraft Anomalies Using LSTMs and Nonparametric Dynamic Thresholding}. In \bibinfo{booktitle}{\emph{Proceedings of the 24th ACM SIGKDD International Conference on Knowledge Discovery \& Data Mining}} (London, United Kingdom) \emph{(\bibinfo{series}{KDD '18})}. \bibinfo{publisher}{Association for Computing Machinery}, \bibinfo{address}{New York, NY, USA}, \bibinfo{pages}{387–395}.
\newblock
\showISBNx{9781450355520}
\href{https://doi.org/10.1145/3219819.3219845}{doi:\nolinkurl{10.1145/3219819.3219845}}


\bibitem[Jebraeilli et~al\mbox{.}(2026)]%
        {jebraeilli2026stqs}
\bibfield{author}{\bibinfo{person}{Anastashia Jebraeilli}, \bibinfo{person}{Chenxu Liu}, \bibinfo{person}{Keyi Yin}, \bibinfo{person}{Samuel Stein}, \bibinfo{person}{Erik Lentz}, \bibinfo{person}{Yufei Ding}, {and} \bibinfo{person}{Ang Li}.} \bibinfo{year}{2026}\natexlab{}.
\newblock \showarticletitle{Stqs: A unified system architecture for spatial temporal quantum sensing}.
\newblock \bibinfo{journal}{\emph{ACM Transactions on Quantum Computing}} \bibinfo{volume}{7}, \bibinfo{number}{2} (\bibinfo{year}{2026}), \bibinfo{pages}{1--34}.
\newblock


\bibitem[Kandala et~al\mbox{.}(2017)]%
        {kandala2017hardware}
\bibfield{author}{\bibinfo{person}{Abhinav Kandala}, \bibinfo{person}{Antonio Mezzacapo}, \bibinfo{person}{Kristan Temme}, \bibinfo{person}{Maika Takita}, \bibinfo{person}{Markus Brink}, \bibinfo{person}{Jerry~M Chow}, {and} \bibinfo{person}{Jay~M Gambetta}.} \bibinfo{year}{2017}\natexlab{}.
\newblock \showarticletitle{Hardware-efficient variational quantum eigensolver for small molecules and quantum magnets}.
\newblock \bibinfo{journal}{\emph{nature}} \bibinfo{volume}{549}, \bibinfo{number}{7671} (\bibinfo{year}{2017}), \bibinfo{pages}{242--246}.
\newblock


\bibitem[Khalid et~al\mbox{.}(2024)]%
        {khalid2024integrated}
\bibfield{author}{\bibinfo{person}{Uman Khalid}, \bibinfo{person}{Muhammad~Shohibul Ulum}, \bibinfo{person}{Moe~Z Win}, {and} \bibinfo{person}{Hyundong Shin}.} \bibinfo{year}{2024}\natexlab{}.
\newblock \showarticletitle{Integrated satellite-ground variational quantum sensing networks}.
\newblock \bibinfo{journal}{\emph{IEEE Communications Magazine}} \bibinfo{volume}{62}, \bibinfo{number}{10} (\bibinfo{year}{2024}), \bibinfo{pages}{20--27}.
\newblock


\bibitem[Lai et~al\mbox{.}(2021)]%
        {lai2021orbitcast}
\bibfield{author}{\bibinfo{person}{Zeqi Lai}, \bibinfo{person}{Qian Wu}, \bibinfo{person}{Hewu Li}, \bibinfo{person}{Mingyang Lv}, {and} \bibinfo{person}{Jianping Wu}.} \bibinfo{year}{2021}\natexlab{}.
\newblock \showarticletitle{Orbitcast: Exploiting mega-constellations for low-latency earth observation}. In \bibinfo{booktitle}{\emph{2021 IEEE 29th International Conference on Network Protocols (ICNP)}}. IEEE, \bibinfo{pages}{1--12}.
\newblock


\bibitem[Lei et~al\mbox{.}(2025)]%
        {lei2025joint}
\bibfield{author}{\bibinfo{person}{Chengjia Lei}, \bibinfo{person}{Shaohua Wu}, \bibinfo{person}{Yi Yang}, \bibinfo{person}{Jiayin Xue}, \bibinfo{person}{Dawei Chen}, \bibinfo{person}{Pengfei Duan}, {and} \bibinfo{person}{Qinyu Zhang}.} \bibinfo{year}{2025}\natexlab{}.
\newblock \showarticletitle{Joint Partitioning, Allocation, and Transmission Optimization for Federated Learning in Satellite Constellations via Multi-Task MARL}.
\newblock \bibinfo{journal}{\emph{IEEE Transactions on Mobile Computing}} (\bibinfo{year}{2025}).
\newblock


\bibitem[Li et~al\mbox{.}(2025b)]%
        {li2025realization}
\bibfield{author}{\bibinfo{person}{Jinting Li}, \bibinfo{person}{Xi Chen}, \bibinfo{person}{Danfang Zhang}, \bibinfo{person}{Wenzhang Wang}, \bibinfo{person}{Yang Zhou}, \bibinfo{person}{Meng He}, \bibinfo{person}{Jie Fang}, \bibinfo{person}{Lin Zhou}, \bibinfo{person}{Chuan He}, \bibinfo{person}{Junjie Jiang}, {et~al\mbox{.}}} \bibinfo{year}{2025}\natexlab{b}.
\newblock \showarticletitle{Realization of a cold atom gyroscope in space}.
\newblock \bibinfo{journal}{\emph{National Science Review}} \bibinfo{volume}{12}, \bibinfo{number}{4} (\bibinfo{year}{2025}), \bibinfo{pages}{nwaf012}.
\newblock


\bibitem[Li et~al\mbox{.}(2025a)]%
        {li2025microsatellite}
\bibfield{author}{\bibinfo{person}{Yang Li}, \bibinfo{person}{Wen-Qi Cai}, \bibinfo{person}{Ji-Gang Ren}, \bibinfo{person}{Chao-Ze Wang}, \bibinfo{person}{Meng Yang}, \bibinfo{person}{Liang Zhang}, \bibinfo{person}{Hui-Ying Wu}, \bibinfo{person}{Liang Chang}, \bibinfo{person}{Jin-Cai Wu}, \bibinfo{person}{Biao Jin}, {et~al\mbox{.}}} \bibinfo{year}{2025}\natexlab{a}.
\newblock \showarticletitle{Microsatellite-based real-time quantum key distribution}.
\newblock \bibinfo{journal}{\emph{Nature}} \bibinfo{volume}{640}, \bibinfo{number}{8057} (\bibinfo{year}{2025}), \bibinfo{pages}{47--54}.
\newblock


\bibitem[Liu et~al\mbox{.}(2025)]%
        {liu2025spaceexit}
\bibfield{author}{\bibinfo{person}{Jiacheng Liu}, \bibinfo{person}{Xiaozhi Zhu}, \bibinfo{person}{Tongqiao Xu}, \bibinfo{person}{Xiaofeng Hou}, {and} \bibinfo{person}{Chao Li}.} \bibinfo{year}{2025}\natexlab{}.
\newblock \showarticletitle{$\{$SpaceExit$\}$: Enabling Efficient Adaptive Computing in Space with Early Exits}. In \bibinfo{booktitle}{\emph{2025 USENIX Annual Technical Conference (USENIX ATC 25)}}. \bibinfo{pages}{1343--1358}.
\newblock


\bibitem[Lu et~al\mbox{.}(2022)]%
        {lu2022micius}
\bibfield{author}{\bibinfo{person}{Chao-Yang Lu}, \bibinfo{person}{Yuan Cao}, \bibinfo{person}{Cheng-Zhi Peng}, {and} \bibinfo{person}{Jian-Wei Pan}.} \bibinfo{year}{2022}\natexlab{}.
\newblock \showarticletitle{Micius quantum experiments in space}.
\newblock \bibinfo{journal}{\emph{Reviews of Modern Physics}} \bibinfo{volume}{94}, \bibinfo{number}{3} (\bibinfo{year}{2022}), \bibinfo{pages}{035001}.
\newblock


\bibitem[Ludmir et~al\mbox{.}(2026a)]%
        {ludmir2026spintune}
\bibfield{author}{\bibinfo{person}{Jason Ludmir}, \bibinfo{person}{Nicholas~S DiBrita}, \bibinfo{person}{Jason Han}, {and} \bibinfo{person}{Tirthak Patel}.} \bibinfo{year}{2026}\natexlab{a}.
\newblock \showarticletitle{SpinTune: Improving the Reliability of Quantum Sensor Networks for Practical Quantum-Classical Utility}.
\newblock \bibinfo{journal}{\emph{arXiv preprint arXiv:2605.04416}} (\bibinfo{year}{2026}).
\newblock


\bibitem[Ludmir et~al\mbox{.}(2026b)]%
        {ludmir2026qufoundrygeneratingdataquantum}
\bibfield{author}{\bibinfo{person}{Jason Ludmir}, \bibinfo{person}{Ian Martin}, \bibinfo{person}{Nicholas~S. DiBrita}, {and} \bibinfo{person}{Tirthak Patel}.} \bibinfo{year}{2026}\natexlab{b}.
\newblock \bibinfo{title}{QuFoundry: Generating Data with Quantum Properties for Quantum Machine Learning Utility}.
\newblock
\showeprint[arxiv]{2509.21622}~[cs.ET]
\urldef\tempurl%
\url{https://arxiv.org/abs/2509.21622}
\showURL{%
\tempurl}


\bibitem[Ludmir et~al\mbox{.}(2025)]%
        {ludmir2025quorum}
\bibfield{author}{\bibinfo{person}{Jason~Zev Ludmir}, \bibinfo{person}{Sophia Rebello}, \bibinfo{person}{Jacob Ruiz}, {and} \bibinfo{person}{Tirthak Patel}.} \bibinfo{year}{2025}\natexlab{}.
\newblock \showarticletitle{Quorum: Zero-Training Unsupervised Anomaly Detection using Quantum Autoencoders}. In \bibinfo{booktitle}{\emph{2025 62nd ACM/IEEE Design Automation Conference (DAC)}}. IEEE, \bibinfo{pages}{1--7}.
\newblock


\bibitem[Malia et~al\mbox{.}(2022)]%
        {malia2022distributed}
\bibfield{author}{\bibinfo{person}{Benjamin~K Malia}, \bibinfo{person}{Yunfan Wu}, \bibinfo{person}{Juli{\'a}n Mart{\'\i}nez-Rinc{\'o}n}, {and} \bibinfo{person}{Mark~A Kasevich}.} \bibinfo{year}{2022}\natexlab{}.
\newblock \showarticletitle{Distributed quantum sensing with mode-entangled spin-squeezed atomic states}.
\newblock \bibinfo{journal}{\emph{Nature}} \bibinfo{volume}{612}, \bibinfo{number}{7941} (\bibinfo{year}{2022}), \bibinfo{pages}{661--665}.
\newblock


\bibitem[Matthiesen et~al\mbox{.}(2024)]%
        {10121575}
\bibfield{author}{\bibinfo{person}{Bho Matthiesen}, \bibinfo{person}{Nasrin Razmi}, \bibinfo{person}{Israel Leyva-Mayorga}, \bibinfo{person}{Armin Dekorsy}, {and} \bibinfo{person}{Petar Popovski}.} \bibinfo{year}{2024}\natexlab{}.
\newblock \showarticletitle{Federated Learning in Satellite Constellations}.
\newblock \bibinfo{journal}{\emph{IEEE Network}} \bibinfo{volume}{38}, \bibinfo{number}{2} (\bibinfo{year}{2024}), \bibinfo{pages}{232--239}.
\newblock
\href{https://doi.org/10.1109/MNET.132.2200504}{doi:\nolinkurl{10.1109/MNET.132.2200504}}


\bibitem[Maule et~al\mbox{.}(2024)]%
        {maule2024fair}
\bibfield{author}{\bibinfo{person}{Ronald Maule}, \bibinfo{person}{Nitish~K Panigrahy}, \bibinfo{person}{Naga~Lakshmi Anipeddi}, \bibinfo{person}{Prajit Dhara}, \bibinfo{person}{Deirdre Kilbane}, \bibinfo{person}{Md~Zakir Hossain}, \bibinfo{person}{Walter~O Krawec}, \bibinfo{person}{Don Towsley}, {and} \bibinfo{person}{Bing Wang}.} \bibinfo{year}{2024}\natexlab{}.
\newblock \showarticletitle{Fair and efficient scheduling strategies for satellite assisted quantum key distribution systems}. In \bibinfo{booktitle}{\emph{2024 IEEE International Conference on Quantum Computing and Engineering (QCE)}}, Vol.~\bibinfo{volume}{1}. IEEE, \bibinfo{pages}{1788--1798}.
\newblock


\bibitem[McMahan et~al\mbox{.}(2017)]%
        {mcmahan2017communication}
\bibfield{author}{\bibinfo{person}{Brendan McMahan}, \bibinfo{person}{Eider Moore}, \bibinfo{person}{Daniel Ramage}, \bibinfo{person}{Seth Hampson}, {and} \bibinfo{person}{Blaise~Aguera y Arcas}.} \bibinfo{year}{2017}\natexlab{}.
\newblock \showarticletitle{Communication-efficient learning of deep networks from decentralized data}. In \bibinfo{booktitle}{\emph{Artificial intelligence and statistics}}. Pmlr, \bibinfo{pages}{1273--1282}.
\newblock


\bibitem[Meesala et~al\mbox{.}(2024a)]%
        {meesala2024quantum}
\bibfield{author}{\bibinfo{person}{Srujan Meesala}, \bibinfo{person}{David Lake}, \bibinfo{person}{Steven Wood}, \bibinfo{person}{Piero Chiappina}, \bibinfo{person}{Changchun Zhong}, \bibinfo{person}{Andrew~D Beyer}, \bibinfo{person}{Matthew~D Shaw}, \bibinfo{person}{Liang Jiang}, {and} \bibinfo{person}{Oskar Painter}.} \bibinfo{year}{2024}\natexlab{a}.
\newblock \showarticletitle{Quantum entanglement between optical and microwave photonic qubits}.
\newblock \bibinfo{journal}{\emph{Physical Review X}} \bibinfo{volume}{14}, \bibinfo{number}{3} (\bibinfo{year}{2024}), \bibinfo{pages}{031055}.
\newblock


\bibitem[Meesala et~al\mbox{.}(2024b)]%
        {meesala2024non}
\bibfield{author}{\bibinfo{person}{Srujan Meesala}, \bibinfo{person}{Steven Wood}, \bibinfo{person}{David Lake}, \bibinfo{person}{Piero Chiappina}, \bibinfo{person}{Changchun Zhong}, \bibinfo{person}{Andrew~D Beyer}, \bibinfo{person}{Matthew~D Shaw}, \bibinfo{person}{Liang Jiang}, {and} \bibinfo{person}{Oskar Painter}.} \bibinfo{year}{2024}\natexlab{b}.
\newblock \showarticletitle{Non-classical microwave--optical photon pair generation with a chip-scale transducer}.
\newblock \bibinfo{journal}{\emph{Nature Physics}} \bibinfo{volume}{20}, \bibinfo{number}{5} (\bibinfo{year}{2024}), \bibinfo{pages}{871--877}.
\newblock


\bibitem[Miroszewski et~al\mbox{.}(2026)]%
        {miroszewski2026quo}
\bibfield{author}{\bibinfo{person}{Artur Miroszewski}, \bibinfo{person}{Jakub Nalepa}, \bibinfo{person}{Agata~M Wijata}, \bibinfo{person}{Jakub Mielczarek}, \bibinfo{person}{Bertrand Le~Saux}, {and} \bibinfo{person}{Alessandro Sebastianelli}.} \bibinfo{year}{2026}\natexlab{}.
\newblock \showarticletitle{Quo Vadis, Quantum Machine Learning?: Quantum kernel methods meet Earth observation}.
\newblock \bibinfo{journal}{\emph{IEEE Geoscience and Remote Sensing Magazine}} (\bibinfo{year}{2026}).
\newblock


\bibitem[Monga et~al\mbox{.}(2023)]%
        {monga2023quant}
\bibfield{author}{\bibinfo{person}{Inder Monga}, \bibinfo{person}{Erhan Saglamyurek}, \bibinfo{person}{Ezra Kissel}, \bibinfo{person}{Hartmut Haffner}, {and} \bibinfo{person}{Wenji Wu}.} \bibinfo{year}{2023}\natexlab{}.
\newblock \showarticletitle{QUANT-NET: A testbed for quantum networking research over deployed fiber}. In \bibinfo{booktitle}{\emph{Proceedings of the 1st Workshop on Quantum Networks and Distributed Quantum Computing}}. \bibinfo{pages}{31--37}.
\newblock


\bibitem[Nielsen and Chuang(2010)]%
        {nielsen2010quantum}
\bibfield{author}{\bibinfo{person}{Michael~A Nielsen} {and} \bibinfo{person}{Isaac~L Chuang}.} \bibinfo{year}{2010}\natexlab{}.
\newblock \bibinfo{booktitle}{\emph{Quantum computation and quantum information}}.
\newblock \bibinfo{publisher}{Cambridge university press}.
\newblock


\bibitem[Panigrahy et~al\mbox{.}(2022)]%
        {panigrahy2022optimal}
\bibfield{author}{\bibinfo{person}{Nitish~K Panigrahy}, \bibinfo{person}{Prajit Dhara}, \bibinfo{person}{Don Towsley}, \bibinfo{person}{Saikat Guha}, {and} \bibinfo{person}{Leandros Tassiulas}.} \bibinfo{year}{2022}\natexlab{}.
\newblock \showarticletitle{Optimal entanglement distribution using satellite based quantum networks}. In \bibinfo{booktitle}{\emph{IEEE INFOCOM 2022-IEEE Conference on Computer Communications Workshops (INFOCOM WKSHPS)}}. IEEE, \bibinfo{pages}{1--6}.
\newblock


\bibitem[Park et~al\mbox{.}(2024)]%
        {park2024dynamic}
\bibfield{author}{\bibinfo{person}{Soohyun Park}, \bibinfo{person}{Soyi Jung}, {and} \bibinfo{person}{Joongheon Kim}.} \bibinfo{year}{2024}\natexlab{}.
\newblock \showarticletitle{Dynamic quantum federated learning for satellite-ground integrated systems using slimmable quantum neural networks}.
\newblock \bibinfo{journal}{\emph{IEEE Access}}  \bibinfo{volume}{12} (\bibinfo{year}{2024}), \bibinfo{pages}{58239--58247}.
\newblock


\bibitem[Park et~al\mbox{.}(2025)]%
        {park2025quantum}
\bibfield{author}{\bibinfo{person}{Soohyun Park}, \bibinfo{person}{Hyunsoo Lee}, \bibinfo{person}{Seok~Bin Son}, \bibinfo{person}{Soyi Jung}, {and} \bibinfo{person}{Joongheon Kim}.} \bibinfo{year}{2025}\natexlab{}.
\newblock \showarticletitle{Quantum federated learning with pole-angle quantum local training and trainable measurement}.
\newblock \bibinfo{journal}{\emph{Neural Networks}}  \bibinfo{volume}{187} (\bibinfo{year}{2025}), \bibinfo{pages}{107301}.
\newblock


\bibitem[Pezze et~al\mbox{.}(2018)]%
        {pezze2018quantum}
\bibfield{author}{\bibinfo{person}{Luca Pezze}, \bibinfo{person}{Augusto Smerzi}, \bibinfo{person}{Markus~K Oberthaler}, \bibinfo{person}{Roman Schmied}, {and} \bibinfo{person}{Philipp Treutlein}.} \bibinfo{year}{2018}\natexlab{}.
\newblock \showarticletitle{Quantum metrology with nonclassical states of atomic ensembles}.
\newblock \bibinfo{journal}{\emph{Reviews of Modern Physics}} \bibinfo{volume}{90}, \bibinfo{number}{3} (\bibinfo{year}{2018}), \bibinfo{pages}{035005}.
\newblock


\bibitem[Pfandzelter et~al\mbox{.}(2025)]%
        {pfandzelter2025trabant}
\bibfield{author}{\bibinfo{person}{Tobias Pfandzelter}, \bibinfo{person}{Nikita Bauer}, \bibinfo{person}{Alexander Leis}, \bibinfo{person}{Corentin Perdrizet}, \bibinfo{person}{Felix Trautwein}, \bibinfo{person}{Trever Schirmer}, \bibinfo{person}{Osama Abboud}, {and} \bibinfo{person}{David Bermbach}.} \bibinfo{year}{2025}\natexlab{}.
\newblock \showarticletitle{Trabant: A Serverless Architecture for Multi-Tenant Orbital Edge Computing}.
\newblock \bibinfo{journal}{\emph{arXiv preprint arXiv:2504.08337}} (\bibinfo{year}{2025}).
\newblock


\bibitem[Pfandzelter and Bermbach(2024)]%
        {pfandzelter2024komet}
\bibfield{author}{\bibinfo{person}{Tobias Pfandzelter} {and} \bibinfo{person}{David Bermbach}.} \bibinfo{year}{2024}\natexlab{}.
\newblock \showarticletitle{Komet: A serverless platform for low-earth orbit edge services}. In \bibinfo{booktitle}{\emph{Proceedings of the 2024 ACM Symposium on Cloud Computing}}. \bibinfo{pages}{866--882}.
\newblock


\bibitem[Pokharel et~al\mbox{.}(2025)]%
        {pokharel2025differentially}
\bibfield{author}{\bibinfo{person}{Atit Pokharel}, \bibinfo{person}{Ratun Rahman}, \bibinfo{person}{Shaba Shaon}, \bibinfo{person}{Thomas Morris}, {and} \bibinfo{person}{Dinh~C Nguyen}.} \bibinfo{year}{2025}\natexlab{}.
\newblock \showarticletitle{Differentially private federated quantum learning via quantum noise}. In \bibinfo{booktitle}{\emph{2025 IEEE International Conference on Quantum Computing and Engineering (QCE)}}, Vol.~\bibinfo{volume}{1}. IEEE, \bibinfo{pages}{1559--1565}.
\newblock


\bibitem[Pompili(2021)]%
        {pompili2021multi}
\bibfield{author}{\bibinfo{person}{Matteo Pompili}.} \bibinfo{year}{2021}\natexlab{}.
\newblock \emph{\bibinfo{title}{Multi-Node Quantum Networks with Diamond Qubits.}}
\newblock \bibinfo{thesistype}{Ph.\,D. Dissertation}. \bibinfo{school}{Delft University of Technology, Netherlands}.
\newblock


\bibitem[Pompili et~al\mbox{.}(2021)]%
        {pompili2021realization}
\bibfield{author}{\bibinfo{person}{Matteo Pompili}, \bibinfo{person}{Sophie~LN Hermans}, \bibinfo{person}{Simon Baier}, \bibinfo{person}{Hans~KC Beukers}, \bibinfo{person}{Peter~C Humphreys}, \bibinfo{person}{Raymond~N Schouten}, \bibinfo{person}{Raymond~FL Vermeulen}, \bibinfo{person}{Marijn~J Tiggelman}, \bibinfo{person}{Laura dos Santos~Martins}, \bibinfo{person}{Bas Dirkse}, {et~al\mbox{.}}} \bibinfo{year}{2021}\natexlab{}.
\newblock \showarticletitle{Realization of a multinode quantum network of remote solid-state qubits}.
\newblock \bibinfo{journal}{\emph{Science}} \bibinfo{volume}{372}, \bibinfo{number}{6539} (\bibinfo{year}{2021}), \bibinfo{pages}{259--264}.
\newblock


\bibitem[Razmi et~al\mbox{.}(2022)]%
        {razmi2022ground}
\bibfield{author}{\bibinfo{person}{Nasrin Razmi}, \bibinfo{person}{Bho Matthiesen}, \bibinfo{person}{Armin Dekorsy}, {and} \bibinfo{person}{Petar Popovski}.} \bibinfo{year}{2022}\natexlab{}.
\newblock \showarticletitle{Ground-assisted federated learning in LEO satellite constellations}.
\newblock \bibinfo{journal}{\emph{IEEE Wireless Communications Letters}} \bibinfo{volume}{11}, \bibinfo{number}{4} (\bibinfo{year}{2022}), \bibinfo{pages}{717--721}.
\newblock


\bibitem[Razmi et~al\mbox{.}(2024)]%
        {10409275}
\bibfield{author}{\bibinfo{person}{Nasrin Razmi}, \bibinfo{person}{Bho Matthiesen}, \bibinfo{person}{Armin Dekorsy}, {and} \bibinfo{person}{Petar Popovski}.} \bibinfo{year}{2024}\natexlab{}.
\newblock \showarticletitle{On-Board Federated Learning for Satellite Clusters With Inter-Satellite Links}.
\newblock \bibinfo{journal}{\emph{IEEE Transactions on Communications}} \bibinfo{volume}{72}, \bibinfo{number}{6} (\bibinfo{year}{2024}), \bibinfo{pages}{3408--3424}.
\newblock
\href{https://doi.org/10.1109/TCOMM.2024.3356429}{doi:\nolinkurl{10.1109/TCOMM.2024.3356429}}


\bibitem[Ren et~al\mbox{.}(2017)]%
        {ren2017ground}
\bibfield{author}{\bibinfo{person}{Ji-Gang Ren}, \bibinfo{person}{Ping Xu}, \bibinfo{person}{Hai-Lin Yong}, \bibinfo{person}{Liang Zhang}, \bibinfo{person}{Sheng-Kai Liao}, \bibinfo{person}{Juan Yin}, \bibinfo{person}{Wei-Yue Liu}, \bibinfo{person}{Wen-Qi Cai}, \bibinfo{person}{Meng Yang}, \bibinfo{person}{Li Li}, {et~al\mbox{.}}} \bibinfo{year}{2017}\natexlab{}.
\newblock \showarticletitle{Ground-to-satellite quantum teleportation}.
\newblock \bibinfo{journal}{\emph{Nature}} \bibinfo{volume}{549}, \bibinfo{number}{7670} (\bibinfo{year}{2017}), \bibinfo{pages}{70--73}.
\newblock


\bibitem[Romero et~al\mbox{.}(2017)]%
        {romero2017quantum}
\bibfield{author}{\bibinfo{person}{Jonathan Romero}, \bibinfo{person}{Jonathan~P Olson}, {and} \bibinfo{person}{Alan Aspuru-Guzik}.} \bibinfo{year}{2017}\natexlab{}.
\newblock \showarticletitle{Quantum autoencoders for efficient compression of quantum data}.
\newblock \bibinfo{journal}{\emph{Quantum Science and Technology}} \bibinfo{volume}{2}, \bibinfo{number}{4} (\bibinfo{year}{2017}), \bibinfo{pages}{045001}.
\newblock


\bibitem[Ruszczak et~al\mbox{.}(2025)]%
        {ruszczak2025ops}
\bibfield{author}{\bibinfo{person}{Bogdan Ruszczak}, \bibinfo{person}{Krzysztof Kotowski}, \bibinfo{person}{David Evans}, {and} \bibinfo{person}{Jakub Nalepa}.} \bibinfo{year}{2025}\natexlab{}.
\newblock \showarticletitle{The OPS-SAT benchmark for detecting anomalies in satellite telemetry}.
\newblock \bibinfo{journal}{\emph{Scientific Data}} \bibinfo{volume}{12}, \bibinfo{number}{1} (\bibinfo{year}{2025}), \bibinfo{pages}{710}.
\newblock


\bibitem[Saito and Rehmsmeier(2015)]%
        {saito2015precision}
\bibfield{author}{\bibinfo{person}{Takaya Saito} {and} \bibinfo{person}{Marc Rehmsmeier}.} \bibinfo{year}{2015}\natexlab{}.
\newblock \showarticletitle{The precision-recall plot is more informative than the ROC plot when evaluating binary classifiers on imbalanced datasets}.
\newblock \bibinfo{journal}{\emph{PloS one}} \bibinfo{volume}{10}, \bibinfo{number}{3} (\bibinfo{year}{2015}), \bibinfo{pages}{e0118432}.
\newblock


\bibitem[Schmidl et~al\mbox{.}(2022)]%
        {10.14778/3538598.3538602}
\bibfield{author}{\bibinfo{person}{Sebastian Schmidl}, \bibinfo{person}{Phillip Wenig}, {and} \bibinfo{person}{Thorsten Papenbrock}.} \bibinfo{year}{2022}\natexlab{}.
\newblock \showarticletitle{Anomaly detection in time series: a comprehensive evaluation}.
\newblock \bibinfo{journal}{\emph{Proc. VLDB Endow.}} \bibinfo{volume}{15}, \bibinfo{number}{9} (\bibinfo{date}{May} \bibinfo{year}{2022}), \bibinfo{pages}{1779–1797}.
\newblock
\showISSN{2150-8097}
\href{https://doi.org/10.14778/3538598.3538602}{doi:\nolinkurl{10.14778/3538598.3538602}}


\bibitem[Schon et~al\mbox{.}(2024)]%
        {schon2024quant}
\bibfield{author}{\bibinfo{person}{Damian Schon}, \bibinfo{person}{Prathwiraj Umesh}, \bibinfo{person}{You-Wei Cheah}, \bibinfo{person}{Se-Young Yu}, \bibinfo{person}{Ezra Kissel}, \bibinfo{person}{Raju Valivarthi}, \bibinfo{person}{Erhan Saglamyurek}, \bibinfo{person}{Lavanya Ramakrishnan}, \bibinfo{person}{Wenji Wu}, \bibinfo{person}{Alp Sipahigil}, {et~al\mbox{.}}} \bibinfo{year}{2024}\natexlab{}.
\newblock \showarticletitle{The QUANT-NET testbed development and preliminary results}. In \bibinfo{booktitle}{\emph{2024 IEEE International Conference on Quantum Computing and Engineering (QCE)}}, Vol.~\bibinfo{volume}{1}. IEEE, \bibinfo{pages}{1799--1808}.
\newblock


\bibitem[Schuld et~al\mbox{.}(2019)]%
        {schuld2019evaluating}
\bibfield{author}{\bibinfo{person}{Maria Schuld}, \bibinfo{person}{Ville Bergholm}, \bibinfo{person}{Christian Gogolin}, \bibinfo{person}{Josh Izaac}, {and} \bibinfo{person}{Nathan Killoran}.} \bibinfo{year}{2019}\natexlab{}.
\newblock \showarticletitle{Evaluating analytic gradients on quantum hardware}.
\newblock \bibinfo{journal}{\emph{Physical Review A}} \bibinfo{volume}{99}, \bibinfo{number}{3} (\bibinfo{year}{2019}), \bibinfo{pages}{032331}.
\newblock


\bibitem[Shen et~al\mbox{.}(2023)]%
        {shen2023hertz}
\bibfield{author}{\bibinfo{person}{Si Shen}, \bibinfo{person}{Chenzhi Yuan}, \bibinfo{person}{Zichang Zhang}, \bibinfo{person}{Hao Yu}, \bibinfo{person}{Ruiming Zhang}, \bibinfo{person}{Chuanrong Yang}, \bibinfo{person}{Hao Li}, \bibinfo{person}{Zhen Wang}, \bibinfo{person}{You Wang}, \bibinfo{person}{Guangwei Deng}, {et~al\mbox{.}}} \bibinfo{year}{2023}\natexlab{}.
\newblock \showarticletitle{Hertz-rate metropolitan quantum teleportation}.
\newblock \bibinfo{journal}{\emph{Light: Science \& Applications}} \bibinfo{volume}{12}, \bibinfo{number}{1} (\bibinfo{year}{2023}), \bibinfo{pages}{115}.
\newblock


\bibitem[Spall(2002)]%
        {spall2002multivariate}
\bibfield{author}{\bibinfo{person}{James~C Spall}.} \bibinfo{year}{2002}\natexlab{}.
\newblock \showarticletitle{Multivariate stochastic approximation using a simultaneous perturbation gradient approximation}.
\newblock \bibinfo{journal}{\emph{IEEE transactions on automatic control}} \bibinfo{volume}{37}, \bibinfo{number}{3} (\bibinfo{year}{2002}), \bibinfo{pages}{332--341}.
\newblock


\bibitem[Stolk et~al\mbox{.}(2024)]%
        {stolk2024metropolitan}
\bibfield{author}{\bibinfo{person}{Arian~J Stolk}, \bibinfo{person}{Kian~L van~der Enden}, \bibinfo{person}{Marie-Christine Slater}, \bibinfo{person}{Ingmar te Raa-Derckx}, \bibinfo{person}{Pieter Botma}, \bibinfo{person}{Joris Van~Rantwijk}, \bibinfo{person}{JJ~Benjamin Biemond}, \bibinfo{person}{Ronald~AJ Hagen}, \bibinfo{person}{Rodolf~W Herfst}, \bibinfo{person}{Wouter~D Koek}, {et~al\mbox{.}}} \bibinfo{year}{2024}\natexlab{}.
\newblock \showarticletitle{Metropolitan-scale heralded entanglement of solid-state qubits}.
\newblock \bibinfo{journal}{\emph{Science advances}} \bibinfo{volume}{10}, \bibinfo{number}{44} (\bibinfo{year}{2024}), \bibinfo{pages}{eadp6442}.
\newblock


\bibitem[Tao et~al\mbox{.}(2024)]%
        {tao2024known}
\bibfield{author}{\bibinfo{person}{Bill Tao}, \bibinfo{person}{Om Chabra}, \bibinfo{person}{Ishani Janveja}, \bibinfo{person}{Indranil Gupta}, {and} \bibinfo{person}{Deepak Vasisht}.} \bibinfo{year}{2024}\natexlab{}.
\newblock \showarticletitle{Known knowns and unknowns: Near-realtime earth observation via query bifurcation in serval}. In \bibinfo{booktitle}{\emph{21st USENIX Symposium on Networked Systems Design and Implementation (NSDI 24)}}. \bibinfo{pages}{809--824}.
\newblock


\bibitem[Tong et~al\mbox{.}(2026)]%
        {tong2026hybrid}
\bibfield{author}{\bibinfo{person}{Ki-Chang Tong}, \bibinfo{person}{Yeong-Gyun Jung}, \bibinfo{person}{Hyeon-Su Yu}, \bibinfo{person}{Young-Hwan Youn}, \bibinfo{person}{Myung-Sun Baek}, \bibinfo{person}{Gyeong-Ho Lee}, {and} \bibinfo{person}{Hyoung-Kyu Song}.} \bibinfo{year}{2026}\natexlab{}.
\newblock \showarticletitle{Hybrid quantum-classical neural networks for low earth orbit satellite communications in 6G}.
\newblock \bibinfo{journal}{\emph{Scientific Reports}} (\bibinfo{year}{2026}).
\newblock


\bibitem[Trinh et~al\mbox{.}(2022)]%
        {trinh2022statistical}
\bibfield{author}{\bibinfo{person}{Phuc~V Trinh}, \bibinfo{person}{Alberto Carrasco-Casado}, \bibinfo{person}{Hideki Takenaka}, \bibinfo{person}{Mikio Fujiwara}, \bibinfo{person}{Mitsuo Kitamura}, \bibinfo{person}{Masahide Sasaki}, {and} \bibinfo{person}{Morio Toyoshima}.} \bibinfo{year}{2022}\natexlab{}.
\newblock \showarticletitle{Statistical verifications and deep-learning predictions for satellite-to-ground quantum atmospheric channels}.
\newblock \bibinfo{journal}{\emph{Communications Physics}} \bibinfo{volume}{5}, \bibinfo{number}{1} (\bibinfo{year}{2022}), \bibinfo{pages}{225}.
\newblock


\bibitem[Valivarthi et~al\mbox{.}(2016)]%
        {valivarthi2016quantum}
\bibfield{author}{\bibinfo{person}{Raju Valivarthi}, \bibinfo{person}{Marcel li~Grimau Puigibert}, \bibinfo{person}{Qiang Zhou}, \bibinfo{person}{Gabriel~H Aguilar}, \bibinfo{person}{Varun~B Verma}, \bibinfo{person}{Francesco Marsili}, \bibinfo{person}{Matthew~D Shaw}, \bibinfo{person}{Sae~Woo Nam}, \bibinfo{person}{Daniel Oblak}, {and} \bibinfo{person}{Wolfgang Tittel}.} \bibinfo{year}{2016}\natexlab{}.
\newblock \showarticletitle{Quantum teleportation across a metropolitan fibre network}.
\newblock \bibinfo{journal}{\emph{Nature Photonics}} \bibinfo{volume}{10}, \bibinfo{number}{10} (\bibinfo{year}{2016}), \bibinfo{pages}{676--680}.
\newblock


\bibitem[Vallado et~al\mbox{.}(2006)]%
        {vallado2006revisiting}
\bibfield{author}{\bibinfo{person}{David Vallado}, \bibinfo{person}{Paul Crawford}, \bibinfo{person}{Ricahrd Hujsak}, {and} \bibinfo{person}{Thomas~Sean Kelso}.} \bibinfo{year}{2006}\natexlab{}.
\newblock \showarticletitle{Revisiting spacetrack report\# 3}. In \bibinfo{booktitle}{\emph{AIAA/AAS astrodynamics specialist conference and exhibit}}. \bibinfo{pages}{6753}.
\newblock


\bibitem[Valley(1980)]%
        {valley1980isoplanatic}
\bibfield{author}{\bibinfo{person}{George~C Valley}.} \bibinfo{year}{1980}\natexlab{}.
\newblock \showarticletitle{Isoplanatic degradation of tilt correction and short-term imaging systems}.
\newblock \bibinfo{journal}{\emph{Applied Optics}} \bibinfo{volume}{19}, \bibinfo{number}{4} (\bibinfo{year}{1980}), \bibinfo{pages}{574--577}.
\newblock


\bibitem[Vasisht et~al\mbox{.}(2021)]%
        {vasisht2021l2d2}
\bibfield{author}{\bibinfo{person}{Deepak Vasisht}, \bibinfo{person}{Jayanth Shenoy}, {and} \bibinfo{person}{Ranveer Chandra}.} \bibinfo{year}{2021}\natexlab{}.
\newblock \showarticletitle{L2D2: Low latency distributed downlink for LEO satellites}. In \bibinfo{booktitle}{\emph{Proceedings of the 2021 ACM SIGCOMM 2021 Conference}}. \bibinfo{pages}{151--164}.
\newblock


\bibitem[Wehner et~al\mbox{.}(2018)]%
        {wehner2018quantum}
\bibfield{author}{\bibinfo{person}{Stephanie Wehner}, \bibinfo{person}{David Elkouss}, {and} \bibinfo{person}{Ronald Hanson}.} \bibinfo{year}{2018}\natexlab{}.
\newblock \showarticletitle{Quantum internet: A vision for the road ahead}.
\newblock \bibinfo{journal}{\emph{Science}} \bibinfo{volume}{362}, \bibinfo{number}{6412} (\bibinfo{year}{2018}), \bibinfo{pages}{eaam9288}.
\newblock


\bibitem[Wengerowsky et~al\mbox{.}(2019)]%
        {wengerowsky2019entanglement}
\bibfield{author}{\bibinfo{person}{S{\"o}ren Wengerowsky}, \bibinfo{person}{Siddarth~Koduru Joshi}, \bibinfo{person}{Fabian Steinlechner}, \bibinfo{person}{Julien~R Zichi}, \bibinfo{person}{Sergiy~M Dobrovolskiy}, \bibinfo{person}{Rene Van~der Molen}, \bibinfo{person}{Johannes~WN Los}, \bibinfo{person}{Val Zwiller}, \bibinfo{person}{Marijn~AM Versteegh}, \bibinfo{person}{Alberto Mura}, {et~al\mbox{.}}} \bibinfo{year}{2019}\natexlab{}.
\newblock \showarticletitle{Entanglement distribution over a 96-km-long submarine optical fiber}.
\newblock \bibinfo{journal}{\emph{Proceedings of the National Academy of Sciences}} \bibinfo{volume}{116}, \bibinfo{number}{14} (\bibinfo{year}{2019}), \bibinfo{pages}{6684--6688}.
\newblock


\bibitem[Wiedmann et~al\mbox{.}(2023)]%
        {wiedmann2023empirical}
\bibfield{author}{\bibinfo{person}{Marco Wiedmann}, \bibinfo{person}{Marc H{\"o}lle}, \bibinfo{person}{Maniraman Periyasamy}, \bibinfo{person}{Nico Meyer}, \bibinfo{person}{Christian Ufrecht}, \bibinfo{person}{Daniel~D Scherer}, \bibinfo{person}{Axel Plinge}, {and} \bibinfo{person}{Christopher Mutschler}.} \bibinfo{year}{2023}\natexlab{}.
\newblock \showarticletitle{An empirical comparison of optimizers for quantum machine learning with spsa-based gradients}. In \bibinfo{booktitle}{\emph{2023 IEEE International Conference on Quantum Computing and Engineering (QCE)}}, Vol.~\bibinfo{volume}{1}. IEEE, \bibinfo{pages}{450--456}.
\newblock


\bibitem[Williams et~al\mbox{.}(2024a)]%
        {williams2024scalable}
\bibfield{author}{\bibinfo{person}{Albert Williams}, \bibinfo{person}{Nitish~K Panigrahy}, \bibinfo{person}{Andrew McGregor}, {and} \bibinfo{person}{Don Towsley}.} \bibinfo{year}{2024}\natexlab{a}.
\newblock \showarticletitle{Scalable scheduling policies for quantum satellite networks}. In \bibinfo{booktitle}{\emph{2024 IEEE International Conference on Quantum Computing and Engineering (QCE)}}, Vol.~\bibinfo{volume}{1}. IEEE, \bibinfo{pages}{1760--1769}.
\newblock


\bibitem[Williams et~al\mbox{.}(2024b)]%
        {williams2024pathfinder}
\bibfield{author}{\bibinfo{person}{Jason~R Williams}, \bibinfo{person}{Charles~A Sackett}, \bibinfo{person}{Holger Ahlers}, \bibinfo{person}{David~C Aveline}, \bibinfo{person}{Patrick Boegel}, \bibinfo{person}{Sofia Botsi}, \bibinfo{person}{Eric Charron}, \bibinfo{person}{Ethan~R Elliott}, \bibinfo{person}{Naceur Gaaloul}, \bibinfo{person}{Enno Giese}, {et~al\mbox{.}}} \bibinfo{year}{2024}\natexlab{b}.
\newblock \showarticletitle{Pathfinder experiments with atom interferometry in the Cold Atom Lab onboard the International Space Station}.
\newblock \bibinfo{journal}{\emph{Nature Communications}} \bibinfo{volume}{15}, \bibinfo{number}{1} (\bibinfo{year}{2024}), \bibinfo{pages}{6414}.
\newblock


\bibitem[Wootters and Zurek(2009)]%
        {wootters2009no}
\bibfield{author}{\bibinfo{person}{William~K Wootters} {and} \bibinfo{person}{Wojciech~H Zurek}.} \bibinfo{year}{2009}\natexlab{}.
\newblock \showarticletitle{The no-cloning theorem}.
\newblock \bibinfo{journal}{\emph{Physics Today}} \bibinfo{volume}{62}, \bibinfo{number}{2} (\bibinfo{year}{2009}), \bibinfo{pages}{76--77}.
\newblock


\bibitem[Wu et~al\mbox{.}(2023)]%
        {wu2023quantum}
\bibfield{author}{\bibinfo{person}{Shuhe Wu}, \bibinfo{person}{Guzhi Bao}, \bibinfo{person}{Jinxian Guo}, \bibinfo{person}{Jun Chen}, \bibinfo{person}{Wei Du}, \bibinfo{person}{Minwei Shi}, \bibinfo{person}{Peiyu Yang}, \bibinfo{person}{Liqing Chen}, {and} \bibinfo{person}{Weiping Zhang}.} \bibinfo{year}{2023}\natexlab{}.
\newblock \showarticletitle{Quantum magnetic gradiometer with entangled twin light beams}.
\newblock \bibinfo{journal}{\emph{Science Advances}} \bibinfo{volume}{9}, \bibinfo{number}{15} (\bibinfo{year}{2023}), \bibinfo{pages}{eadg1760}.
\newblock


\bibitem[Wurtz et~al\mbox{.}(2023)]%
        {wurtz2023aquila}
\bibfield{author}{\bibinfo{person}{Jonathan Wurtz}, \bibinfo{person}{Alexei Bylinskii}, \bibinfo{person}{Boris Braverman}, \bibinfo{person}{Jesse Amato-Grill}, \bibinfo{person}{Sergio~H Cantu}, \bibinfo{person}{Florian Huber}, \bibinfo{person}{Alexander Lukin}, \bibinfo{person}{Fangli Liu}, \bibinfo{person}{Phillip Weinberg}, \bibinfo{person}{John Long}, {et~al\mbox{.}}} \bibinfo{year}{2023}\natexlab{}.
\newblock \showarticletitle{Aquila: Quera's 256-qubit neutral-atom quantum computer}.
\newblock \bibinfo{journal}{\emph{arXiv preprint arXiv:2306.11727}} (\bibinfo{year}{2023}).
\newblock


\bibitem[Yang et~al\mbox{.}(2025)]%
        {yang2025clock}
\bibfield{author}{\bibinfo{person}{YA Yang}, \bibinfo{person}{Maya Miklos}, \bibinfo{person}{Yee~Ming Tso}, \bibinfo{person}{Stella Kraus}, \bibinfo{person}{Joonseok Hur}, {and} \bibinfo{person}{Jun Ye}.} \bibinfo{year}{2025}\natexlab{}.
\newblock \showarticletitle{Clock precision beyond the Standard Quantum Limit at 10-18 level}.
\newblock \bibinfo{journal}{\emph{Physical Review Letters}} \bibinfo{volume}{135}, \bibinfo{number}{19} (\bibinfo{year}{2025}), \bibinfo{pages}{193202}.
\newblock


\bibitem[Yin et~al\mbox{.}(2017)]%
        {yin2017satellite}
\bibfield{author}{\bibinfo{person}{Juan Yin}, \bibinfo{person}{Yuan Cao}, \bibinfo{person}{Yu-Huai Li}, \bibinfo{person}{Sheng-Kai Liao}, \bibinfo{person}{Liang Zhang}, \bibinfo{person}{Ji-Gang Ren}, \bibinfo{person}{Wen-Qi Cai}, \bibinfo{person}{Wei-Yue Liu}, \bibinfo{person}{Bo Li}, \bibinfo{person}{Hui Dai}, {et~al\mbox{.}}} \bibinfo{year}{2017}\natexlab{}.
\newblock \showarticletitle{Satellite-based entanglement distribution over 1200 kilometers}.
\newblock \bibinfo{journal}{\emph{Science}} \bibinfo{volume}{356}, \bibinfo{number}{6343} (\bibinfo{year}{2017}), \bibinfo{pages}{1140--1144}.
\newblock


\bibitem[Zhang et~al\mbox{.}(2024)]%
        {zhang2024resource}
\bibfield{author}{\bibinfo{person}{Qiyang Zhang}, \bibinfo{person}{Xin Yuan}, \bibinfo{person}{Ruolin Xing}, \bibinfo{person}{Yiran Zhang}, \bibinfo{person}{Zimu Zheng}, \bibinfo{person}{Xiao Ma}, \bibinfo{person}{Mengwei Xu}, \bibinfo{person}{Schahram Dustdar}, {and} \bibinfo{person}{Shangguang Wang}.} \bibinfo{year}{2024}\natexlab{}.
\newblock \showarticletitle{Resource-efficient in-orbit detection of earth objects}. In \bibinfo{booktitle}{\emph{IEEE INFOCOM 2024-IEEE Conference on Computer Communications}}. IEEE, \bibinfo{pages}{551--560}.
\newblock


\end{thebibliography}

\clearpage
\appendix
\DeclareUrlCommand\fpath{\urlstyle{tt}}
\newcommand{\zenodolatest}{\href{https://doi.org/10.5281/zenodo.22909579}{\nolinkurl{doi.org/10.5281/zenodo.22909579}}}
\newcommand{\zenodoeval}{\href{https://doi.org/10.5281/zenodo.23162894}{\nolinkurl{doi.org/10.5281/zenodo.23162894}}}
\newtcblisting{cmd}{listing only, colback=black!5, colframe=black!5,
  boxrule=0pt, arc=1mm, left=1.5mm, right=1mm, top=0.7mm, bottom=0.7mm,
  boxsep=0pt, before skip=4pt, after skip=4pt,
  listing options={basicstyle=\footnotesize\ttfamily, columns=fullflexible,
  keepspaces=true, breaklines=false, aboveskip=0pt, belowskip=0pt}}
\setlength{\emergencystretch}{2.5em}
\raggedbottom

\section{Artifact Appendix}
\label{sec:artifact}

\subsection{Abstract}
\label{sec:artifact:abstract}

Our artifact consists of the trained models of the paper and the code that
evaluates and trains them. We claim that \sol{} improves detection quality,
latency, throughput, and QPU utilization over the isolated, centralized,
and FedAverage baselines and outperforms Quorum (Sec.~\ref{sec:eval}),
holds its gain on other QPU backends and under the sensor noise
(Sec.~\ref{sec:eval:robustness}), and scales with the sensing rate and the
fleet size (Sec.~\ref{sec:eval:scaling}). The artifact contains the trained
autoencoders of every method, the inputs they are evaluated on, and one
script per figure that regenerates it and prints every value next to the
paper's. The latest version of the archive also provides the real-hardware path, which needs an IBM Quantum account and AWS Braket
access. By default, the circuits run in Qiskit Aer as noisy simulations under the backends' calibration data, 
and the sensor noise comes from the shipped spectroscopy, so no quantum hardware account is needed in that case.

\subsection{Artifact check-list (meta-information)}
\label{sec:artifact:checklist}

\begin{itemize}[leftmargin=*]
  \item \textbf{Program:} The \sol{} package: the autoencoder bank, router,
  thresholds, channel and QPU noise models, sensor channel, serving model,
  federated training, and one script per figure.
  \item \textbf{Model:} The trained autoencoders of every method on OPS-SAT
  and SMAP with their training logs; the scores of Quorum's 50 members.
  \item \textbf{Data set:} OPS-SAT and NASA SMAP as published, the two event
  streams, the IBM calibration snapshots, and the Aquila sensor noise.
  \item \textbf{Hardware:} In simulation, an NVIDIA GPU (our run times are
  for four A100s). On real hardware, an IBM Quantum account and AWS Braket
  access to QuEra Aquila.
  \item \textbf{How much disk space required (approximately)?:} 32\,GB
  \item \textbf{How much time is needed to prepare workflow
  (approximately)?:} $\sim$15 minutes
\item \textbf{How much time is needed to complete experiments
  (approximately)?:} 1\,h\,15\,min for the main figures on four A100s,
  about 3 hours more for Quorum and Figs.~\ref{fig:sensor} and
  \ref{fig:scaling}; 15 to 20 hours per dataset to retrain the models. On
  real hardware, the time depends on the backends' queues.
  \item \textbf{Publicly available?:} Yes: \zenodolatest{} (latest
  version; simulation and real hardware)
  \item \textbf{Code licenses (if publicly available)?:} MIT
\end{itemize}
\subsection{Description}
\label{sec:artifact:description}

\subsubsection{How to access}
Download \fpath{EntangledEdge-Artifact.zip} from \zenodolatest{} and
unpack it. Every module runs from the root of the archive as
\texttt{python -m} \fpath{entanglededge.<module>} and lists its options
with \texttt{--help}.

\subsubsection{Hardware dependencies}
In simulation, an NVIDIA GPU with a CUDA~12 driver; a scoring job needs
1\,GB of GPU memory, 3\,GB of host memory, and 4 CPU threads, and the
sensor spectroscopy on the Bloqade emulator needs 16 CPU cores. On real
hardware, an IBM Quantum account with access to at least one backend (by
default the scripts use the backends of the calibration snapshots;
\texttt{--ibm-backends} names others) and AWS Braket access to QuEra
Aquila.

\subsubsection{Software dependencies}
Linux x86-64, an NVIDIA driver with CUDA~12 support, and Miniconda;
\fpath{setup.sh} installs everything else (\fpath{requirements.txt}). On
real hardware, an IBM Quantum token and AWS credentials; the sensor
spectroscopy on the Bloqade emulator also needs Julia with Bloqade.

\subsubsection{Data sets}
All inputs ship in \fpath{data/}. It holds the OPS-SAT and SMAP datasets of
Sec.~\ref{sec:method:dataset}, with the raw SMAP telemetry in
\fpath{smap_raw/}. \fpath{events/} holds the event streams of both
datasets, \fpath{backends/} the calibration data of the four IBM backends
as Aer noise models, and \fpath{sensor/} the Aquila noise spectroscopy and
the sensor channel fitted to it.

\subsubsection{Models}
The artifact ships every trained model so that \sol{} can be tested
without the long training jobs (\fpath{models/}). The scores of the 50
members of the Quorum ensemble (Sec.~\ref{sec:method:quorum}) are shipped
at \fpath{models/opssat/Quorum-50members/}.

\subsection{Installation}
\label{sec:artifact:install}

From the root of the unpacked archive:
\begin{cmd}
./setup.sh
conda activate entanglededge
python reproduce/kick_the_tires.py
\end{cmd}
The smoke test takes about 7 minutes on one GPU and ends with
\texttt{All stages ran.} For real hardware, add the IBM Quantum token and
the AWS credentials as in README Sections 9 and 10.

\subsection{Experiment workflow}
\label{sec:artifact:workflow}

\noindent\textbf{1. Training.} To train the models of a dataset
(\texttt{--dataset smap} for SMAP), one trainer per model with the
arguments of the paper's runs, in Aer under the stations' calibration
data:
\begin{cmd}
python reproduce/train_flagship.py \
    --dataset opssat --gpus 0,1,2,3
\end{cmd}
or on the IBM backends:
\begin{cmd}
python reproduce/train_flagship.py \
    --dataset opssat --qpu ibm --ibm-shots 512
\end{cmd}
Either writes the models with their training logs to
\fpath{results/<dataset>/}; \fpath{configs/<dataset>.yaml} selects them
for the deployment below in place of the shipped ones in \fpath{models/}.

\vspace{1mm}

\noindent\textbf{2. Deployment.} The trained models serve the held-out
event streams of Sec.~\ref{sec:method:eval}: every event is scored, routed,
thresholded, and run through the serving model. This one run yields
Figs.~\ref{fig:flagship_system} to \ref{fig:quorum} and
\ref{fig:ablation}. In Aer, \fpath{reproduce_all.sh} runs the figure
scripts, and the backend sweep of point 3, in one go:
\begin{cmd}
./reproduce_all.sh
\end{cmd}
On the IBM backends, only the scoring script changes, with
\texttt{--qpu ibm}; the other four run no circuits:
\begin{cmd}
python reproduce/figure10.py --qpu ibm
python reproduce/figure9.py
python reproduce/figure11.py
python reproduce/figure12.py
python reproduce/figure16.py
\end{cmd}
where \texttt{--ibm-max-events N} limits the scoring to the first $N$
events of each stream and a rerun continues from the last finished job.
Quorum's 50 members score the same OPS-SAT stream; the artifact ships the
scores of that run, and
\begin{cmd}
python reproduce/quorum_ensemble.py --gpus 0,1,2,3
python reproduce/figure12.py \
    --members-dir outputs/figure12/members
\end{cmd}
produces them again in Aer. Every figure lands as PDF in \fpath{figures/},
and \fpath{comparison_with_paper.txt} lists every value of the run next to
the paper's.

\vspace{1mm}

\noindent\textbf{3. Robustness.} The deployment is run again under other
QPU backends (Fig.~\ref{fig:backends}) and under the sensor noise
(Fig.~\ref{fig:sensor}). The backend sweep, in Aer under each
configuration's calibration data, or with each station on the backend of
its configuration:
\begin{cmd}
python reproduce/figure13.py --gpus 0,1,2,3
python reproduce/figure13.py --qpu ibm
\end{cmd}
The sensor noise is the channel fitted to the noise spectroscopy of one
Aquila atom (Sec.~\ref{sec:method:sensor}), on the noisy Bloqade emulator
of Aquila:
\begin{cmd}
python -m entanglededge.sensor.spectroscopy \
    --backend bloqade --shots 10000 \
    --out outputs/sensor/noise_spectroscopy.npz
\end{cmd}
or measured on the device through AWS Braket (\texttt{--backend aquila
--shots 1000}); the spectral density, the channel, and the deployment
under it follow:
\begin{cmd}
python -m entanglededge.sensor.nsd \
    --spectroscopy \
    outputs/sensor/noise_spectroscopy.npz \
    --out outputs/sensor/nsd_fit.json
python -m entanglededge.sensor.channel \
    --nsd outputs/sensor/nsd_fit.json \
    --out outputs/sensor/sensor_channel.npz
python reproduce/figure15.py --gpus 0,1,2,3 \
    --sensor-channel \
    outputs/sensor/sensor_channel.npz
\end{cmd}
Without \texttt{--sensor-channel}, \fpath{figure15.py} uses the shipped
channel in \fpath{data/sensor/}; with \texttt{--qpu ibm}, its circuits run
on the IBM backends.

\vspace{1mm}

\noindent\textbf{4. Scaling.} The deployment is run on the thirteen event
streams of Sec.~\ref{sec:eval:scaling}, four sensing rates and nine fleet
sizes, which the script generates:
\begin{cmd}
python reproduce/figure17.py --gpus 0,1,2,3
\end{cmd}
\texttt{--qpu ibm} runs their circuits on the IBM backends, as in point 2.

\subsection{Evaluation and expected results}
\label{sec:artifact:results}

\noindent\textbf{1. Training.} Retrained models are statistically, not
bitwise, equivalent to the shipped ones, since the shot sampling is not
seeded; run through the deployment, they are expected to reproduce the
ranking of the methods within the tolerance of point 2.

\vspace{1mm}

\noindent\textbf{2. Deployment.} The figures are expected to show the
paper's trends, with \sol{} ahead of every baseline and of Quorum on F1,
and \fpath{comparison_with_paper.txt} lists every value next to the
paper's. In Aer, the metrics are expected within a few percent of the
paper's values. On the IBM backends, the outputs are the same files, with
the backend, the shots, and the job IDs recorded in each run's manifest;
this path was not part of the artifact evaluation, and we state no
tolerance for it.

\vspace{1mm}

\noindent\textbf{3. Robustness.} \sol{} is expected to keep the largest F1
improvement over Isolated in every backend configuration
(Fig.~\ref{fig:backends}; the artifact uses \fpath{ibm_fez} in place of
Torino, whose calibration data are not available) and a positive
improvement in nearly every chunk as the sensor distortion grows
(Fig.~\ref{fig:sensor}). A spectroscopy measured on Aquila enters through
the same fit and channel as the shipped one.

\vspace{1mm}

\noindent\textbf{4. Scaling.} Utilization and throughput are expected to
rise with the sensing rate, throughput with the fleet size, while
per-event latency and the F1 improvement over Isolated stay roughly
constant and positive; the improvement comes out higher than the paper's,
which used an earlier version of the event pipeline.

\end{document}